\documentclass[draft]{agujournal2019}
\usepackage{url}
\usepackage{amsmath,amssymb}
\usepackage{tikz}
\definecolor{orcidlogocol}{HTML}{A6CE39}
\newcommand{\orcidlogo}{\raisebox{-0.15ex}{\tikz[baseline]{\fill[orcidlogocol]
  (0,0.5ex) circle (0.75ex); \node[white,inner sep=0,font=\fontsize{3.6}{3.6}\selectfont\bfseries\sffamily]
  at (0,0.5ex) {iD};}}}
\newcommand{\orcidlink}[1]{\pdfstartlink attr{/Border[0 0 0]}
  user{/Subtype/Link/A<</Type/Action/S/URI/URI(https://orcid.org/#1)>>}%
  \orcidlogo\pdfendlink}
\usepackage{placeins}
\draftfalse

\newcommand{\zlat}{\mathbf{z}}
\newcommand{\mlat}{\mathbf{m}}

\newcommand{\Dec}{\mathcal{D}}

\journalname{Journal of Advances in Modeling Earth Systems (JAMES)}

\begin{document}

\title{WRF-AeroMELD: Advecting a Learned Representation of
Particle-Resolved Aerosol}

\authors{Jeffrey~H.~Curtis\affil{1}\,\orcidlink{0000-0002-1447-2127},
Wenhan~Tang\affil{1}\,\orcidlink{0009-0009-5076-036X},
Nicole~Riemer\affil{1}\,\orcidlink{0000-0002-3220-3457},
and Matthew~West\affil{2}\,\orcidlink{0000-0002-7605-0050}}

\affiliation{1}{Department of Climate, Meteorology \& Atmospheric Sciences,
   University of Illinois Urbana-Champaign, Urbana, Illinois, USA}
\affiliation{2}{Department of Mechanical Science and Engineering, University
of Illinois Urbana-Champaign, Urbana, Illinois, USA}

\correspondingauthor{Nicole Riemer}{nriemer@illinois.edu}

\begin{keypoints}
\item A learned aerosol representation that resolves mixing state is advected
      in a weather model as ten ordinary tracers.
\item In an idealized 12-hour simulation its average error is
  4 percent compared with 12 to 19 percent for sectional and modal
  schemes.
\item Most sectional and modal schemes spread dust from a few particles over a
      whole bin or mode and overpredict the frozen fraction by a factor of two or three.
\end{keypoints}

\begin{abstract}
AeroMELD was recently developed to represent a particle-resolved
aerosol population, including its mixing state, by its number
concentration and nine learned latent variables.  Each latent variable
is a number-weighted mean of a learned per-particle
feature. Multiplying each variable by the aerosol number mixing ratio
$N$ therefore gives a quantity that can be mixed by dry-air mass. In
WRF-AeroMELD, the Weather Research and Forecasting model (WRF)
transports $N$ and these nine products as ten tracers using its
standard monotonic advection scheme. A learned decoder diagnoses size
distributions, cloud condensation nuclei (CCN) activity, immersion
freezing, and optical properties from the transported state. We
evaluate WRF-AeroMELD in an idealized 12-h simulation over central
California with three prescribed particle populations whose properties
remain fixed during transport. We compute reference diagnostics
directly from the source particles using the number mixing ratio
contributed by each population. We compare WRF-AeroMELD with the
sectional schemes MOSAIC and TOMAS and the modal schemes MAM4 and
MATRIX under the same transport. Averaged over six aerosol properties,
the error of WRF-AeroMELD is 4\%, compared with
12--19\% for the sectional and modal schemes, which use 25--362
tracers. For all five schemes, the overall error after transport is
within about one percentage point of the representation error. The
differences among schemes are largest for immersion freezing: MOSAIC,
TOMAS and MAM4 spread dust from a few percent of the urban particles
over all particles in a bin or mode and overestimate the frozen
fraction two- to threefold. This work enables a mixing-state-resolving
aerosol representation to be transported in regional models with ten
tracers.
\end{abstract}

\section*{Plain Language Summary}
Atmospheric particles differ from one another in what they are made
of. This variation, called the aerosol mixing state, affects how
particles form cloud droplets and ice crystals and how they scatter
and absorb sunlight. Weather and climate models cannot afford to track
individual particles, so they usually assume that all particles of a
similar size have the same composition. AeroMELD is a machine-learning
method that describes a population of particles with ten numbers,
constructed so that when two air masses mix, their numbers are
averaged. We add AeroMELD to a weather model, which moves the ten
numbers with the wind like any other pollutant, and reconstruct the
properties of the particles from them. We test this in a 12-hour
simulation over central California using three groups of particles
whose properties stay fixed as the wind carries them. By tracking how
many particles from each group reach each location, we can calculate
the properties of the local mixture directly. In areas where particles
from different sources mix, the four comparison schemes have average
errors about three to five times as large as AeroMELD's across the six
aerosol properties. These schemes carry 25 to 362 quantities with the
wind, compared with ten for AeroMELD. The largest differences are for
ice formation, which in this simulation depends on dust contained in
only a few particles.  Three of the four comparison schemes spread
this dust over more particles. As a result, the fraction of particles
they predict will freeze is two to three times the reference value.

\section{Introduction}
\label{sec:introduction}

Atmospheric aerosol particles affect climate by acting as cloud condensation
nuclei (CCN) and ice-nucleating particles and by scattering and absorbing
radiation \cite{IPCC2021,Poeschl2005,Mcfiggans2006}. These effects depend not
only on the size distribution and bulk composition of the aerosol but also on
its mixing state \cite{Riemer2019}, that is, on how the chemical species are
distributed among the individual particles \cite{Winkler1973,Riemer2013}. In a
fully external mixture each particle contains a single species, and in a fully
internal mixture all particles have the same composition. The atmospheric
aerosol lies between these limits, and its mixing state evolves over the
particles' atmospheric lifetime. Mixing state influences CCN activity
\cite{Fierce2013,Ching2017}, absorption by black carbon
\cite{Bond2006,Fierce2016,Yao2022} and immersion freezing
\cite{Hoose2012,Tang2026}.

Three-dimensional chemical transport models and Earth system models represent
the aerosol with bulk, modal or sectional schemes
\cite{Whitby1997,Binkowski2003,Vignati2004,Stier2005,Bauer2008,Liu2016,Kaiser2014,Jacobson2002,Zaveri2008}.
Modal schemes describe the size distribution as a sum of lognormal modes, and
sectional schemes divide it into size bins. Within each mode or bin the
particles are assumed to be internally mixed \cite{Fierce2017}, so these
schemes prescribe the mixing state rather than resolve it. The mixing state can
be partly resolved by adding modes or by binning in composition as well as in
size \cite{Ching2016,Matsui2013}. However, the number of bins grows
exponentially with the number of composition dimensions, so in practice only
two or three dimensions are resolved, whereas the atmospheric aerosol contains
tens to hundreds of species. Approximating the mixing state in this way affects
the estimated optical and CCN properties \cite{Bauer2013,Fierce2017,Fierce2024},
and aerosol models disagree widely in their predictions \cite{Mann2014}.

Particle-resolved models avoid these assumptions by representing the
aerosol as a large number of computational particles, each with its
own composition \cite{Riemer2009,Zaveri2010}. The particle-resolved
model PartMC--MOSAIC has been coupled to the Weather Research and
Forecasting model (WRF), first in a single column \cite{Curtis2017}
and then in three dimensions \cite{Curtis2019}, and stochastic
algorithms have been developed to advect the particles
\cite{Curtis2024}. However, particle-resolved simulations are
expensive in three dimensions, because each grid cell requires the
transport of roughly $10^3$--$10^4$ computational particles
\cite{Gasparik2020}, each carrying tens of species.

Machine learning offers a way to reduce this cost. Learned representations have
been used to estimate mixing-state metrics \cite{Hughes2018,Zheng2021,Shen2024},
to emulate aerosol processes \cite{Harder2022,Wang2022,Ferracina2025} and to
generate binned aerosol distributions \cite{saleh2025generative}.
Low-dimensional latent representations have also been learned for cloud
microphysics \cite{Lamb2024,DeJong2026}, including latent representations of
droplet size distributions that can be advected \cite{Huang2025}. To be used in
a transport model, however, an aerosol representation must handle the mixing of
air masses. Advection mixes the air of neighboring grid cells, so the
representation of a mixture must follow from the representations of its parts,
which is not the case for a general autoencoder. AeroMELD
\cite{SalehAeroMELD2026} was designed to have this property. Its encoder
computes learned features for each particle and averages them over the
population, so the latent vector of a mixture of populations is the
number-weighted average of their latent vectors. Coagulation has since been
modeled directly in the same latent space \cite{TangAeroMELDCoag2026}.

AeroMELD has not yet been used in a transport model, and it has not been
compared with the sectional and modal schemes that such models use. Exact
mixing in latent space does not by itself ensure that the advected aerosol is
accurate, because both the host model's advection and the decoder can introduce
errors. Host models advect tracers with nonlinear schemes whose flux limiters
prevent new extrema from forming at sharp fronts
\cite{Zalesak1979,Skamarock2006,Wang2009}. Because each tracer is limited
separately, an advected latent vector is in general not an exact mixture of the
populations that contributed to it. In addition, the decoder must reproduce the
diagnostics of a mixture from its averaged latent vector, which a nonlinear
decoder does not do unless it is trained to.

In this paper we present WRF-AeroMELD, which advects the AeroMELD
state in WRF as ten tracers with WRF's standard monotonic advection
scheme, and we compare it with four sectional and modal schemes: the
sectional schemes MOSAIC \cite{Zaveri2008}, with the 8 bins used in
WRF-Chem, and TOMAS \cite{Adams2002,Lee2013}, with 40 bins, and the
modal schemes MAM4 \cite{Liu2016} and MATRIX \cite{Bauer2008}. All
five representations are advected in the same way and evaluated
against the same particle-resolved reference. The test case is an
idealized 12-h simulation over central California with three
prescribed particle populations. An urban population initially
occupies a box around San Francisco Bay, a background population
initially fills the rest of the domain and enters at the lateral
boundaries, and a fixed inland source emits particles throughout the
run. Because particle properties remain fixed, we can compute
reference diagnostics in each grid cell directly from the source
particles once their respective number mixing ratios are known. This
reference exists only because no process changes the particles, so we
use it to evaluate the representations, but it could not be used as a
full aerosol model itself. We evaluate six diagnostics, namely the
number and speciated mass size distributions, the CCN spectrum, the
frozen fraction, and the scattering and absorption coefficients, and
we separate the error of each representation into a transport error
and a representation error. The appendices examine how the results
depend on the transport scheme, by comparing two other ways of
advecting the latent state, and on the decoder, by comparing four
other decoders. Section~\ref{sec:theory} describes WRF-AeroMELD, the
sectional and modal schemes, the particle-resolved reference and the
error measure. Section~\ref{sec:methods} describes the experiments,
Section~\ref{sec:results} presents the results, and
Sections~\ref{sec:discussion} and~\ref{sec:conclusions} discuss them
and summarize our conclusions.

\section{WRF-AeroMELD and the Comparison Schemes}
\label{sec:theory}

\begin{figure}
\noindent\includegraphics[width=\textwidth]{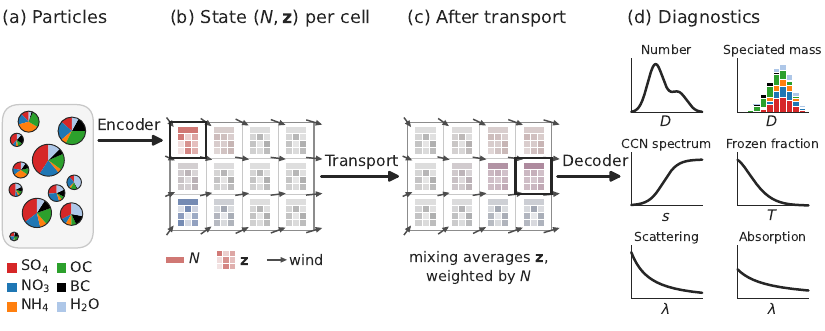}
\caption{Overview of WRF-AeroMELD. (a)~A population of particles, each
  drawn as a pie of its species masses, is mapped by the encoder to
  the state of one grid cell: the number mixing ratio $N$ (bar) and
  the nine components of the latent vector $\zlat$ ($3{\times}3$
  block). (b)~Every cell of the host model's grid holds its own
  $(N,\zlat)$; colors show the mixture of the background (gray), urban
  (red) and emission (blue) populations, shading shows the value of
  each entry, and arrows at the cell vertices show the wind. (c)~The
  host model transports $N$ and $N\zlat$ with the same wind, so the
  plumes move downwind, converge and mix. Under ideal mixing, the
  latent vector of mixed air is the number-weighted average of the
  source vectors. (d)~The decoder turns $(N,\zlat)$ of any cell
  (outlined) into the diagnostics: number and speciated mass size
  distributions, CCN spectrum, frozen fraction, and scattering and
  absorption coefficients. The drawing is illustrative, not model
  output.}
\label{fig:schematic}
\end{figure}

WRF-AeroMELD represents the aerosol in every grid cell of a host model, here
WRF, by the AeroMELD state \cite{SalehAeroMELD2026}, which consists of a number
mixing ratio and a short latent vector. Figure~\ref{fig:schematic} illustrates
the approach. An encoder maps a particle population to the state, the host
model advects the state with the wind, and a decoder computes the diagnostics
of any grid cell from its state. Section~\ref{sec:theory-aeromeld} describes
the representation, its encoder and its decoder, and
Section~\ref{sec:theory-transport} describes how the state is advected.
Section~\ref{sec:theory-baselines} describes the sectional and modal schemes
with which we compare WRF-AeroMELD, and Sections~\ref{sec:theory-reference}
and~\ref{sec:theory-errors} describe the particle-resolved reference and how
the errors are measured against it.

\subsection{The Representation}
\label{sec:theory-aeromeld}
\label{sec:theory-decoders}

AeroMELD describes the aerosol in a grid cell by its number mixing ratio $N$
and a latent vector $\zlat$ that describes the population independently of
its amount \cite{SalehAeroMELD2026}. The encoder computes a short list of
learned features for each particle from its species masses, and the latent
vector is the number-weighted average of these features over the particles:
\begin{equation}
  N=\sum_p w_p,\qquad
  \zlat=\frac{1}{N}\sum_p w_p\,\boldsymbol\phi(x_p),
  \label{eq:encoder}
\end{equation}
where $w_p$ is the number weight of particle $p$, $x_p$ its species masses,
and $\boldsymbol\phi$ a small neural network. The latent vector has nine
components, so the state is ten numbers per cell
(Figure~\ref{fig:schematic}a), and the encoding works for any number of
particles, in any order. $N$ is the number of particles per kilogram of dry
air, the unit in which WRF-Chem advects aerosol number. A mixing ratio is
conserved by WRF's flux-form advection, does not change when air is compressed
or expanded, and is averaged by air mass when air masses mix. The number
concentration per cubic meter is $N$ times the density of dry air.

Because the latent vector is an average over particles, it is also averaged
when air masses mix. When air parcels $k$ mix, each making up a fraction $a_k$
of the mixture's air mass, the number mixing ratio of the mixture is the
air-mass-weighted average of their number mixing ratios, and its latent vector
is exactly the number-weighted average of their latent vectors:
\begin{equation}
  N_{\mathrm{mix}}=\sum_k a_k N_k,\qquad
  \zlat_{\mathrm{mix}}=\frac{1}{N_{\mathrm{mix}}}\sum_k a_k N_k\zlat_k.
  \label{eq:mix}
\end{equation}
Both $N$ and the number-weighted latent vector $N\zlat$ are therefore averaged
by air mass when air masses mix, and a host model can advect them as ten
ordinary tracers.

A decoder $\Dec$ computes the diagnostics of a grid cell from its state,
$\mathbf y=\Dec(N,\zlat)$ (Figure~\ref{fig:schematic}d): the number and
speciated mass size distributions against dry diameter $D$ (the diameter of
a particle without its water; the speciated mass includes the water), the CCN spectrum
(the \emph{activated fraction}, the fraction of particles that activate as
cloud condensation nuclei, against supersaturation $s$), the frozen fraction
(the fraction of particles frozen, against temperature), and the scattering
and absorption coefficients (the \emph{optics}) against wavelength.
Because $\zlat$ describes the population independently of its amount, the
diagnostics are of two kinds. The activated and frozen fractions are fractions
of the particles, so they do not change when the number of particles doubles
and the population is otherwise unchanged. They therefore depend on $\zlat$
alone. The size distributions and the optics instead measure the amount of
aerosol in the air, so they double when the number of particles doubles. Each
of them is $N$ times an average per particle, such as the fraction of particles
in each size bin or the mass of each species per particle, and this average
depends on $\zlat$ alone. We write $\mathbf q(\zlat)$ for the per-particle
diagnostics, the two fractions and these averages per particle, so that
$\Dec(N,\zlat)$ is $\mathbf q(\zlat)$ for the fractions and $N\mathbf q(\zlat)$
for the rest.
Like $\boldsymbol\phi$, $\mathbf q$ is a small neural network, a multilayer
perceptron (MLP) whose outputs are all six diagnostics.

Like its latent vector, the per-particle diagnostics of a mixture are the
number-weighted averages of those of its parts. A perfect decoder would
therefore map the averaged latent vector of a mixture to the average of the
diagnostics of its parts, but a nonlinear decoder does not do so in general.
Because transport mixes air in every grid cell, the decoder of WRF-AeroMELD is
trained on mixtures of up to 16 populations as well as on single populations.
It learns to reproduce the diagnostics of each mixture from its averaged latent
vector \cite{zhang2018mixup,SalehAeroMELD2026}, so that the diagnostics it
computes for a mixture are close to the mixture of the diagnostics
(\ref{app:training}). \ref{app:decoders} compares this decoder with four
alternatives.

\subsection{Transport}
\label{sec:theory-transport}

The host model advects $N$ and the nine components of $N\zlat$ as ordinary
chemical tracers, here in WRF-Chem \cite{Grell2005}, with WRF's standard tracer
advection, in the same way that WRF-Chem advects the tracers of its sectional
and modal schemes. Emissions add the $N$ and $N\zlat$ of the emitted
population, and inflow at the lateral boundaries adds those of the background
population. WRF's tracer advection is in flux form and monotonic. At each cell
face it combines a high-order flux with a first-order upwind flux, using a
limiter coefficient chosen so that the tracer develops no new extrema
\cite{Zalesak1979,Skamarock2006,Wang2009}. Because each tracer has its own
limiter coefficients, the advected latent vector is not necessarily the exact
number-weighted average of Equation~\eqref{eq:mix}. \ref{app:schemes} compares
this approach with two transport schemes that keep mixing exact and shows that
WRF's standard tracer advection gives the smallest transport error.

\subsection{The Sectional and Modal Schemes}
\label{sec:theory-baselines}

We compare WRF-AeroMELD with four schemes used in three-dimensional models
(Table~\ref{tab:schemes}): two sectional schemes, MOSAIC with the 8 bins of
WRF-Chem \cite{Zaveri2008,Fast2006} and TOMAS with the 40 bins of GEOS-Chem
\cite{Adams2002,Lee2013,Trivitayanurak2008},
and two modal schemes, MAM4 \cite{Liu2016} and MATRIX \cite{Bauer2008}, the
latter in its most detailed configuration. In every grid cell each scheme
tracks the number of particles and the mass of each of its species in each
size bin or mode, so these tracers are averaged by air mass when air masses
mix, as $N$ and $N\zlat$ are. All particles in a bin or mode are assumed to
have the same composition. This is the assumption of internal mixing, which
WRF-AeroMELD does not make. The sectional schemes resolve size with fixed bins.
The modal schemes describe each mode by a lognormal distribution of fixed
width. MAM4 has three modes defined by size and one for freshly emitted primary
carbon. MATRIX has 16 populations defined by composition and, within some
classes, by size. These are sulfate, dust and black carbon, each with and
without a coating or by size, sea salt, organic matter, and mixtures of black
carbon with dust, with organic matter, and with everything else.

\begin{table}
\caption{The representations compared: how they resolve size and composition,
and the tracers each advects per grid cell (number and species masses per bin
or mode; $N$ and the nine components of $N\zlat$ for WRF-AeroMELD).}
\label{tab:schemes}
\centering
\begin{tabular}{llll}
\hline
Representation & Size & Composition & Tracers \\
\hline
WRF-AeroMELD (this work) & learned & learned, not assumed & 10 \\
MOSAIC 8 & 8 bins in dry diameter & one per bin, 15 species & 128 \\
TOMAS-40 & 40 bins in dry mass & one per bin, 8 species & 362 \\
MAM4 & 4 lognormal modes & one per mode, 3--7 species & 25 \\
MATRIX & 16 lognormal populations & one per population, 5 species & 53 \\
\hline
\end{tabular}
\end{table}

We use the same three PartMC source populations for all five aerosol
representations. For the sectional schemes, we assign particles to bins
by dry diameter in MOSAIC and by dry mass per particle in TOMAS. For
the modal schemes, we assign particles to modes according to size and
composition. We sum the particle number in each bin or mode and group
the 15 PartMC species according to the chemical species each scheme
carries. In all five representations, water is a passive constituent.

Each scheme transports its own tracers using the same WRF advection
scheme as WRF-AeroMELD, with a separate flux limiter for each tracer.
After each transport step, we apply the adjustments of each scheme that
act on any state, as its host models do, such as corrections to particle
number and transfers between bins or modes. Adjustments that belong to
aerosol processes, such as the aging of fresh carbon, are omitted, since
no process acts in this test case. To calculate aerosol properties, we use
representative particles that span the sizes within each bin or mode
and share its composition. We apply the same particle-level diagnostic
calculations to these particles as to the particle-resolved reference
(Section~\ref{sec:theory-reference}), so the comparison measures the
representations and not differences in the diagnostic calculations.
\ref{app:baselines} describes the species groupings, adjustments, and diagnostic calculations
for each scheme.

\subsection{The Particle-Resolved Reference}
\label{sec:theory-reference}

Evaluating the representations requires the true diagnostics in every
grid cell. In the test case of this paper these are known
(Section~\ref{sec:methods-case}). The model transports the populations
supplied by the initial conditions, lateral boundary inflow, and inland
emissions without changing their size or composition, so the number
fraction of each source population in a grid cell can be determined by
tracking each source separately. We advect the number of each source
population as a separate tracer with WRF's monotonic advection
(\emph{per-source transport}). This gives the number $N^{\mathrm{PS}}$
in every grid cell and its source fractions $f^{\mathrm{PS}}_s$, the
fractions of its particles that come from each source~$s$. The
\emph{particle-resolved reference} combines these with the
per-particle diagnostics $\mathbf q^{\mathrm{PR}}_s$ of each source
population, computed from its particles in the same way as the
training targets (Section~\ref{sec:methods-ae}):
\begin{equation}
  \mathbf y^{\mathrm{PR}}=N^{\mathrm{PS}}\sum_s f^{\mathrm{PS}}_s\,\mathbf q^{\mathrm{PR}}_s,
  \label{eq:exact}
\end{equation}
without the factor $N^{\mathrm{PS}}$ for the activated and frozen
fractions.  For the number mixing ratios supplied by per-source
transport, Equation~(\ref{eq:exact}) gives the exact diagnostics of
the source-particle mixture. The number mixing ratios themselves are
subject to numerical advection error. Per-source transport has the
numerical errors of WRF's tracer advection, which every tracer shares,
so a comparison with it isolates the errors of representing the
aerosol by latent variables or by bins and modes. The reference is
used only for testing. It requires the sources to be known, and it is
not available once coagulation, condensation or chemistry change the
particles. In that case particle-resolved simulations must serve as
the reference. Per-source transport also gives, in every grid cell,
the WRF-AeroMELD state obtained by combining the source latent vectors
in the proportions given by per-source transport. This state consists
of the number $N^{\mathrm{PS}}$ and the exact mixture of the latent
vectors $\zlat_s$ of the sources, $\zlat^{\mathrm{PS}}=\sum_s
f^{\mathrm{PS}}_s\zlat_s$. The exact mixture of the sources of each
sectional or modal scheme is obtained in the same way.

\subsection{Measuring Transport and Representation Errors}
\label{sec:theory-errors}

We evaluate each aerosol representation using six diagnostics: the
number size distribution, the speciated mass size distribution (the
mass of every species in every size bin), the CCN spectrum, the frozen
fraction, and the scattering and absorption coefficients. For a
diagnostic $\hat{\mathbf y}$ and its reference $\mathbf y$, each a
vector of $n$ values over the size bins, species, supersaturations,
temperatures or wavelengths of the diagnostic, the error is
\begin{equation}
  e(\hat{\mathbf y},\mathbf y)=
  \begin{cases}
    \dfrac{\lVert\hat{\mathbf y}-\mathbf y\rVert}{\lVert\hat{\mathbf y}\rVert+\lVert\mathbf y\rVert}
      & \text{size distributions and CCN spectrum},\\[2ex]
    \dfrac12\Big[\dfrac1n\sum_{j=1}^{n}\big(\ln\hat y_j-\ln y_j\big)^2\Big]^{1/2}
      & \text{frozen fraction and optics},
  \end{cases}
  \label{eq:error}
\end{equation}
with the Euclidean norm. The first form is the symmetric error used for
AeroMELD \cite{SalehAeroMELD2026}. Because the frozen fraction and the optics
span orders of magnitude, for them we use half the root-mean-square logarithmic
ratio. Both forms treat overestimates and underestimates alike, and for small
errors both are equal to half the relative error, so an error of 1\% has the
same meaning for every diagnostic. As on the test set (\ref{app:training}), the size
distributions are compared in 15 bins of dry diameter up to 10~$\mu$m, the
speciated mass with its water, the
CCN spectrum at supersaturations of 0.1--0.6\%, the frozen fraction from
$-25$ to $-10\,^{\circ}$C, and the optics at eight wavelengths. The
\emph{overall error} is the mean of $e$ over the six diagnostics.

In the test case, errors are computed in every grid cell and averaged over the
\emph{mixed} cells, in which at least two source populations each contribute
at least 5\% of the number (Section~\ref{sec:methods-metrics}). Outside the
mixed cells a grid cell contains essentially one source population, and its
error is the representation's error for that population alone, which is the
same everywhere. These cells cover more than three quarters of the surface and
would dominate an average over all cells, hiding the errors due to transport
and mixing. We report errors in percent.

The \emph{error} of a representation compares the diagnostics of its
transported state with the particle-resolved reference,
$e\big(\Dec(N,\zlat),\,\mathbf y^{\mathrm{PR}}\big)$, where for a sectional or
modal scheme $(N,\zlat)$ stands for its tracers and $\Dec$ for the
diagnostics of its representative particles. To see where this error comes
from, we ask two questions. How accurately does each representation reproduce
the particle-resolved diagnostics when it is given the reference particle
number and source fractions? How much do its predicted diagnostics change
when the model transports the representation's own tracers?

To answer the first question, we use per-source transport
(Section~\ref{sec:theory-reference}) to construct the exact mixture
of the sources for each representation. For WRF-AeroMELD, this gives the state
$(N^{\mathrm{PS}},\zlat^{\mathrm{PS}})$. For a sectional or modal scheme,
we combine the source bin or mode tracers in the same
proportions. Comparing the diagnostics of this mixture with the
particle-resolved reference gives the \emph{representation error}.
Comparing the diagnostics of the transported state with those of the
same mixture gives the \emph{transport error}. Because $e$ obeys the
triangle inequality,
\begin{equation}
  \underbrace{e\big(\Dec(N,\zlat),\,\mathbf y^{\mathrm{PR}}\big)}_{\text{error}}
  \le\underbrace{e\big(\Dec(N,\zlat),\,\Dec(N^{\mathrm{PS}},\zlat^{\mathrm{PS}})\big)}_{\text{transport error}}
  +\underbrace{e\big(\Dec(N^{\mathrm{PS}},\zlat^{\mathrm{PS}}),\,\mathbf y^{\mathrm{PR}}\big)}_{\text{representation error}}.
  \label{eq:budget}
\end{equation}
for every representation. This bound holds in each grid cell and therefore
also on average, although the two errors do not add exactly. For
WRF-AeroMELD, the representation error is the decoder error. For the
sectional and modal schemes, it reflects how the bins or modes represent
particle size and composition. Their transport error also includes the
effects of any adjustments applied after advection.
\ref{app:attribution} examines each error in more detail.

\section{Experiments}
\label{sec:methods}

\subsection{Training Data, the Decoder and Test Mixtures}
\label{sec:methods-ae}

The decoder is trained on the library of PartMC box-model scenarios of
\citeA{Gasparik2020}: 1000 scenarios with 25 snapshots each, every snapshot a
population of particles with 15 species. As in \citeA{SalehAeroMELD2026},
each population was reduced to 1000 particles, drawn from the simulated
particles. These 1000-particle populations are the particle-resolved
populations of this paper, used both for training and as the sources of the
test case. The diagnostics that the decoder learns to reproduce are computed
from the particles. The CCN spectrum is computed with $\kappa$-K\"ohler theory
\cite{Riemer2010}, scattering and absorption with Mie theory and a core--shell
treatment of black carbon and dust \cite{Yao2022}, and immersion freezing with
an active-site parameterization
\cite{Hoose2012,niemand2012particle,Schill2020,Tang2026}. The three scenarios
used as source populations in the test case (Section~\ref{sec:methods-case}) are
held out of training. The decoder is trained with the averaging encoder of
Equation~\eqref{eq:encoder}, on single populations and on mixtures of up to 16
of them, each mixture encoded as the number-weighted average of its
populations' latent vectors (\ref{app:training}). We train the decoder ten
times with different random seeds, and each seed holds out a different set of
5000 test populations.

We test the decoder with the error of Equation~\eqref{eq:error} on single test
populations and on mixtures of $K$ test populations with random weights, from
$K=2$ up to the whole test set ($K=5000$). The sectional and modal schemes,
which require no training, are tested on the same populations and mixtures with
the same ten splits. Each test population is projected onto the bins or modes
of the scheme from its particles, the mixtures are formed by mixing the
scheme's tracers, and the diagnostics are computed from representative
particles (Section~\ref{sec:theory-baselines}).

\subsection{The Test Case}
\label{sec:methods-case}
\label{sec:methods-ic}

The test case is 17~June~2010, 00--12~UTC, over central California during the
Carbonaceous Aerosols and Radiative Effects Study, which sampled the
Sacramento urban plume \cite{Zaveri2012}. WRF version~4.7.1
\cite{Skamarock2019} runs on a grid of $169\times159$ cells at 4~km with 39
levels, driven by North American Mesoscale analyses (\ref{app:transport}).
The simulation uses no boundary-layer scheme and no explicit diffusion, so the
aerosol is moved by resolved advection only, and the aerosol has no feedback on
radiation or clouds.

\begin{figure}
\noindent\includegraphics[width=\textwidth]{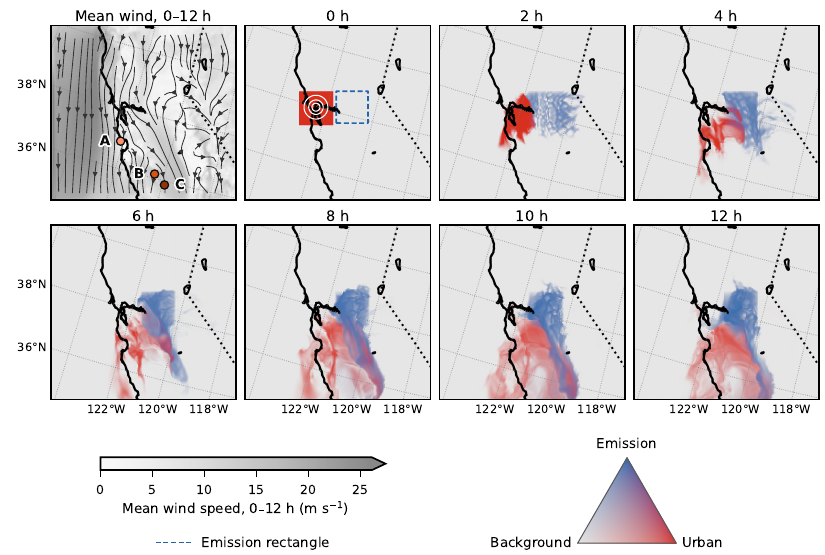}
\caption{The test case on the lowest model level of the whole WRF domain
($169\times159$ cells at 4~km, $676\times636$~km). Top left: the mean wind
over the 12-h run (00--12~UTC on 17~June~2010): its speed, with
streamlines and the three sampling points A, B and~C used in
Section~\ref{sec:results-case}. Other panels: the composition of the aerosol by
source population every 2~h, each cell colored by the number fractions of the
background (gray), urban (red) and emission (blue) populations, mixed as in the
triangle key. At 0~h the urban population alone fills a box around the San
Francisco Bay, so the box is uniformly urban in composition; its number is not
uniform but Gaussian, rising from the background's $10^{9}$~kg$^{-1}$ at the
edge of the box to $10^{10}$~kg$^{-1}$ at its center (white contours at 2, 5
and $8\times10^{9}$~kg$^{-1}$). The background fills the rest of the domain,
and the emission population is emitted throughout the run from a fixed
rectangle further inland (the emission rectangle, dashed blue outline in the
0-h panel), where the blue plume
starts.}
\label{fig:setup}
\end{figure}

\begin{figure}
\noindent\includegraphics[width=\textwidth]{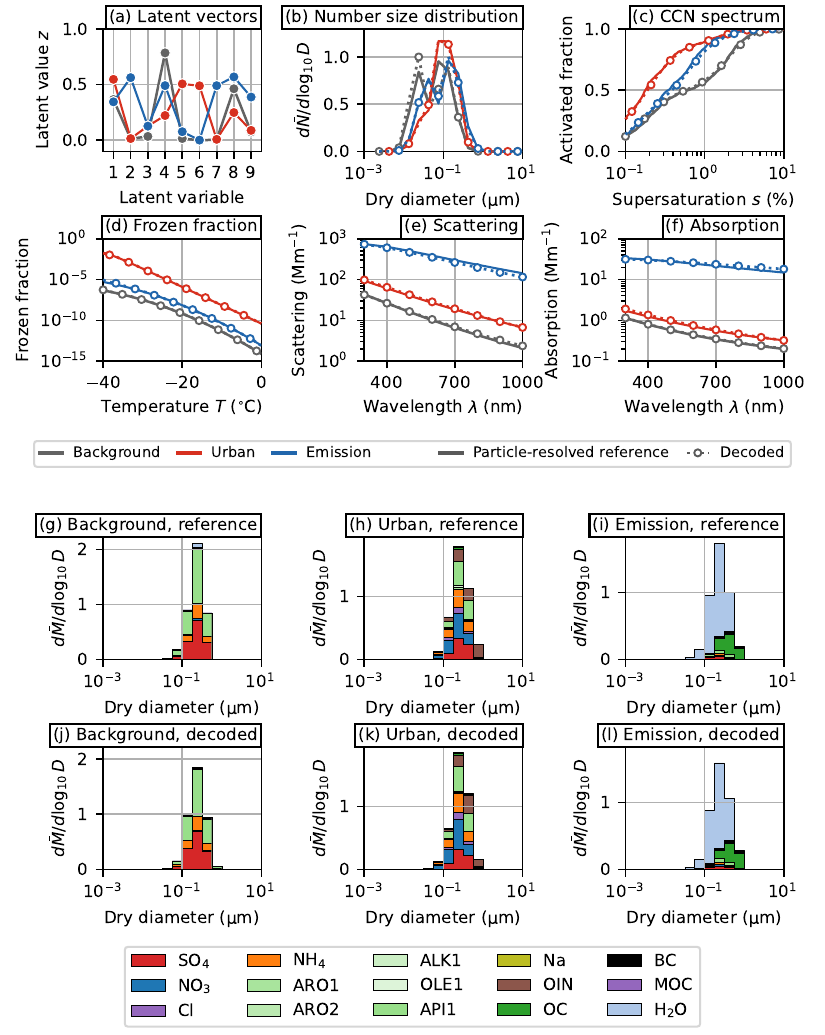}
\caption{The three source populations (background, urban and emission; gray, red
and blue, as in every figure), and their round trip through the encoder and
the decoder of WRF-AeroMELD (representative seed).
Solid lines are the particle-resolved populations, the reference; dotted
lines with open circles their reconstruction after encoding and decoding.
(a)~The nine latent variables of each population. (b)~Number size
distributions against dry diameter, $d\bar N/d\log_{10} D$ with $\bar
N=N/N_{\mathrm{tot}}$, so that each integrates to one. (c)~CCN spectra (activated fraction against
supersaturation). (d)~Frozen fraction against temperature. (e)~Scattering
and (f)~absorption coefficients against wavelength, per cubic meter at the
decoder's nominal number concentration. (g--i)~Speciated mass size
distributions of the particle-resolved populations and (j--l)~of their
reconstructions, stacked by species with water included, $d\bar M/d\log_{10} D$ with $\bar
M=M/M_{\mathrm{tot}}$ and $M_{\mathrm{tot}}$ the reconstruction's total mass,
so that each reconstruction integrates to one.}
\label{fig:populations}
\end{figure}

The aerosol is built from three particle-resolved source populations, three
scenarios of the library of \citeA{Gasparik2020} (561, 635 and 812, each 12~h
into its run), which are held out of training
(Figure~\ref{fig:setup}). A \emph{background} population initially fills the
domain outside the urban region and flows in at the lateral boundaries. An
\emph{urban} population initially fills a box around the San Francisco Bay, with
a number mixing ratio that rises from the background value at the edge of the
box to ten times that value at its center. An \emph{emission} population is
emitted at $5\times10^{7}$~m$^{-2}$~s$^{-1}$ from a fixed \emph{emission
rectangle} farther inland, over Sacramento and the Sierra Nevada foothills.
Coagulation, gas--particle chemistry, deposition and nucleation are switched
off, so no process changes the composition of a particle, and the aerosol is
only emitted and advected. Every grid cell therefore contains a mixture of the
three source populations, mixed by transport across the interfaces between the
initially separate regions, and the particle-resolved reference of
Section~\ref{sec:theory-reference} gives its exact diagnostics from the
particles of the sources. The test case is deliberately idealized. It is
designed as a controlled test of the transport of the latent state, not as a
simulation of the real atmosphere. The urban plume reaches the southern
boundary after about 6~h (Figure~\ref{fig:setup}).
Figure~\ref{fig:populations} shows the three source populations and their
reconstruction by the decoder.

Because the aerosol is passive, we run WRF once and save the air-mass fluxes
used by its tracer advection. Every representation is then advected offline
with these fluxes by a re-implementation of WRF's tracer advection, with the
same initial and boundary conditions and emissions. This allows us to advect
the sectional and modal schemes, with up to 362 tracers, and the alternatives
tested in the appendices for all ten training seeds. When WRF-AeroMELD is run
inside WRF, it agrees with the offline transport throughout the 12-h
simulation, and its overall error against the reference after 12~h is the
same to three digits (\ref{app:transport}).

\subsection{Evaluation}
\label{sec:methods-metrics}

The errors are evaluated on the lowest model level at 12~UTC, the end of the
run, over the mixed cells of per-source transport
(Section~\ref{sec:theory-errors}), which are 5965 of the 26\,871 surface cells.
For maps of a single supersaturation, temperature or wavelength we show the
ratio of the value from WRF-AeroMELD to that of the reference. Because the
decoder is trained with ten seeds, the comparison with the sectional and modal
schemes, the maps and the results at individual points use one
\emph{representative} seed. This is the seed whose errors in the six
diagnostics are jointly closest to the medians over the ten seeds, measured by
the sum over the six diagnostics of $\lvert\ln(e/\mathrm{median})\rvert$. The
sectional and modal schemes require no training, and each has a single run.

\section{Results}
\label{sec:results}

\subsection{Comparison with Sectional and Modal Schemes}
\label{sec:results-comparison}

\begin{figure}
\noindent\includegraphics[width=\textwidth]{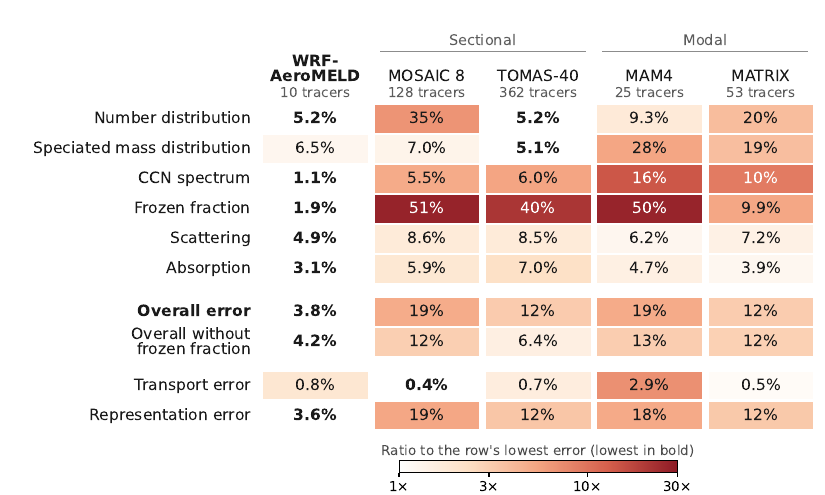}
\caption{Mean error over the 5965 mixed cells after 12~h of
  WRF-AeroMELD and the sectional and modal schemes, each transported
  by WRF's standard tracer advection: per diagnostic, overall (mean
  over the six diagnostics), and overall without the frozen fraction;
  and the transport and representation errors defined in
  Equation~(\eqref{eq:budget}). Each row's lowest error is bold, and
  every cell is shaded by its ratio to it.}
\label{fig:scorecard}
\end{figure}

Figure~\ref{fig:scorecard} compares the errors of the five representations
after 12~h. WRF-AeroMELD has an overall error of 3.8\%, compared with
11.8--19.0\% for the sectional and modal schemes, and its overall error ranges
from 3.3 to 4.1\% across the ten training seeds. It has the lowest error in four
of the six diagnostics, lower than that of the best sectional or modal scheme by
factors from 1.2 for absorption and 1.3 for scattering to 5 for the CCN spectrum
and the frozen fraction. In the number size distribution, WRF-AeroMELD and TOMAS-40 have
the same error (5.2\%), and in the speciated mass size distribution TOMAS-40 has
a lower error (5.1\%, compared with 6.5\%).

For all five schemes, the overall error after transport is within
about one percentage point of the representation error
(Figure~\ref{fig:scorecard}). For WRF-AeroMELD, for example, these
errors are 3.8\% and 3.6\%, respectively. The transport error defined
in Equation~(\ref{eq:budget}) compares the diagnostics of the
transported state directly with those of the exact mixture of the sources and is
0.8\% for WRF-AeroMELD. It is below 1\% for all schemes except MAM4,
for which it reaches 2.9\% because the size adjustment increases
particle number. The fresh carbon particles in the emission population
have a dry diameter of about 220~nm. Assigning them to MAM4's
primary-carbon mode pushes its number-median dry diameter above the
100~nm upper bound. The size adjustment then increases the number of
particles in the mode, bringing the median diameter back toward that
bound. When this carbon is instead projected onto the accumulation
mode, the adjustment remains inactive, and MAM4's overall error
increases by only about half a percentage point
(\ref{app:baselines}). The differences therefore lie in the
representations.  The representation errors of the sectional and modal
schemes are 3 to 5 times that of WRF-AeroMELD.

The largest differences are in the frozen fraction, and they result from the
treatment of the mixing state. Nearly all ice-active particles are in the urban
population, and only 4\% of its particles contain dust. A sectional or modal
scheme assigns one composition to all particles in a bin or mode and therefore
spreads this dust over every particle in the bin or mode. Mixing spreads it
further, over the background and emission particles in the same bin or mode.
Because the ice-nucleating surface of a particle increases as its dust volume to
the power 2/3, a concave function, spreading the same dust over more particles
increases the frozen fraction. In the mixed cells the increase is a factor of
about 2.7 to 2.8 for MOSAIC~8 and MAM4, of which three fifths to two thirds
comes from the bins or modes of each source and the rest from mixing. Additional bins reduce this
error only slowly. The bins of TOMAS-40, which span a factor of 1.26 in diameter
compared with 2 for MOSAIC~8, spread the dust over fewer particles, but even
with 400 bins the frozen-fraction error for the urban population alone is 21\%.
WRF-AeroMELD keeps the dust on the particles that contain it, as the reference
does. MATRIX, whose populations are defined by composition, does so in part.
The urban particles that contain dust, together with the few that contain sea
salt, make up 4.5\% of the urban particles and are assigned to its mixed
population, so their dust is not spread over the other particles. Without the
frozen fraction, the overall errors of the sectional and modal schemes are 1.5
to 3.1 times that of WRF-AeroMELD, and TOMAS-40 comes closest.

The errors in the other diagnostics reflect the structure of each scheme. The
sectional schemes bin by dry size, which is the coordinate of the size
distributions, so for these two diagnostics their errors are limited only by
their bins. The lowest bin edge of MOSAIC~8 is at 39~nm. Between 13 and 36\% of
the particles of the source populations are smaller than this, and the scheme
moves them up to the edge, so its error in the number distribution is 35\%.
Little mass lies below this edge, and its error in the speciated mass
distribution is 7.0\%. TOMAS-40, with 40 bins, has errors of about 5\% in both
distributions, the same as that of WRF-AeroMELD in the number distribution and
lower in the speciated mass distribution. Measuring how far number and
mass are displaced in size, rather than comparing the distributions bin by bin,
leads to the same conclusion. Both sectional schemes place the mass at least as
well as WRF-AeroMELD, and TOMAS-40 also places the number better
(\ref{app:baselines}). MAM4's error in the speciated mass distribution comes
from the single composition of each mode and from its small number of species,
with nitrate and ammonium lumped into sulfate. Its error in the CCN spectrum
comes from its fixed mode widths, from the single composition of each mode and
from the particles created by its size adjustment. The errors of MATRIX in the
number distribution and the CCN spectrum come from its populations defined by
composition. Nearly every particle of the three sources contains some black
carbon and organic matter, so 86--96\% of the particles of each source are
assigned to MATRIX's population of black carbon with organic matter. This
population is a single lognormal mode with a geometric standard deviation of
1.8, whatever the size of its particles. MATRIX's single organic species, which
includes both primary and secondary organic matter, adds to its error in the
speciated mass distribution.

\begin{figure}
\centering
\includegraphics[width=0.6\textwidth]{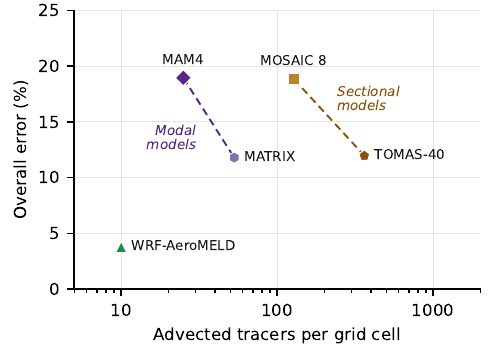}
\caption{Overall error after 12~h over the mixed cells (mean over the six
diagnostics) against the number of tracers each representation advects per
grid cell, on a logarithmic axis: the number and species masses per bin or
mode for the sectional and modal schemes, $N$ and the nine components of
$N\zlat$ for WRF-AeroMELD. Dashed lines join the two schemes of each family.}
\label{fig:tracers}
\end{figure}

Figure~\ref{fig:tracers} shows the overall errors against the number of tracers
that each representation advects per grid cell. Each tracer is advected, stored
and communicated separately, so the memory and the cost of transport grow in
proportion to the number of tracers. Within each family, more tracers give a
lower error. MATRIX, with twice as many tracers as MAM4, has 38\% less
error, and TOMAS-40, with nearly three times as many tracers as MOSAIC~8, has
36\% less. Across the families, the sectional schemes need more tracers
than the modal schemes to reach the same error, and WRF-AeroMELD has a lower
error than either with fewer tracers. The families differ in how they divide
the particles. The sectional schemes divide them by size alone, and the modal
schemes also divide them by composition, slightly in MAM4's primary-carbon mode
and more in MATRIX's populations. WRF-AeroMELD does not divide the particles
into groups at all.

\begin{figure}
\centering
\includegraphics[width=0.6\textwidth]{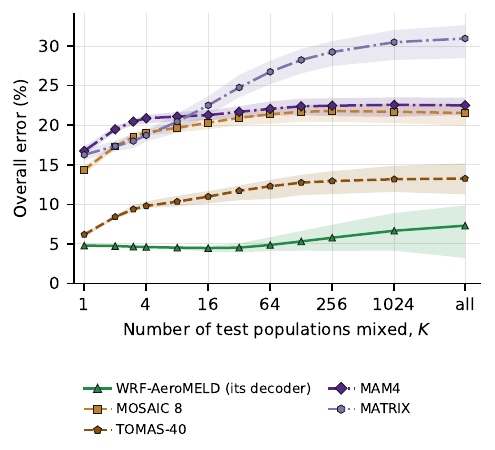}
\caption{Overall error (mean over the six diagnostics) on mixtures of $K$
test populations with Dirichlet weights, against the true mixture, from single
populations ($K=1$) to the whole test set ($K=5000$): the decoder of
WRF-AeroMELD (mean over ten training seeds, lines, with the range, bands) and
the sectional and modal schemes (mean and range over the same ten test
splits).}
\label{fig:kcurves-baselines}
\end{figure}

The test set shows the same difference for any number of mixed populations
(Figure~\ref{fig:kcurves-baselines}). On single test populations TOMAS-40 comes
closest to the decoder. As populations are mixed, the errors of the sectional
and modal schemes increase, because in every bin or mode of a mixture the
particles of many populations share one composition. The error of the decoder,
in contrast, remains nearly constant up to the 16 populations on which it is
trained. For mixtures of a few populations, most of the increase for the
sectional and modal schemes is in the frozen fraction, except for MATRIX. For
mixtures of many populations, most of it is in absorption, because the black
carbon of a few populations is spread over the particles of all of them. The test case mixes at most three sources, and the
difference widens for mixtures of more populations, up to 16. Beyond 16
populations the error of the decoder also increases, in the speciated mass
distribution, but its mean remains below those of the sectional and modal
schemes.

\subsection{WRF-AeroMELD in the Test Case}
\label{sec:results-case}

\begin{figure}
\noindent\includegraphics[width=\textwidth]{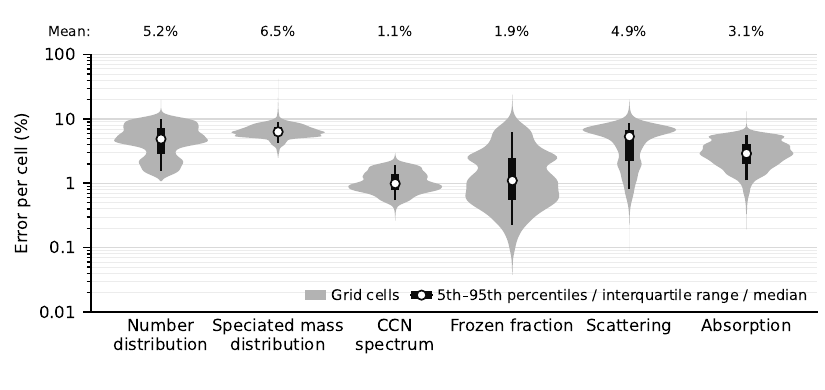}
\caption{Distribution over the 5965 mixed cells after 12~h of the error of
each diagnostic (Equation~\eqref{eq:error}) of WRF-AeroMELD
against the particle-resolved reference. Violins are kernel densities of
$\log_{10} e$; bars mark the 5th--95th percentiles (thin) and the
interquartile range (thick), and white dots the median. The number above
each violin is the mean over the cells, the error quoted in the text.}
\label{fig:summary-error}
\end{figure}

The remaining figures compare WRF-AeroMELD at its representative seed with the
particle-resolved reference after 12~h. This comparison tests the whole chain
from the source particles through encoding, transport and decoding.
Figure~\ref{fig:summary-error} shows the distribution of the error of each
diagnostic over the mixed cells. The distributions are narrow. In five of the
six diagnostics, 95\% of the cells have errors less than twice the mean, and in
the frozen fraction less than 3.4 times the mean, so the mean is not determined
by a few poorly decoded cells.

\begin{figure}
\centering
\includegraphics[width=0.98\textwidth]{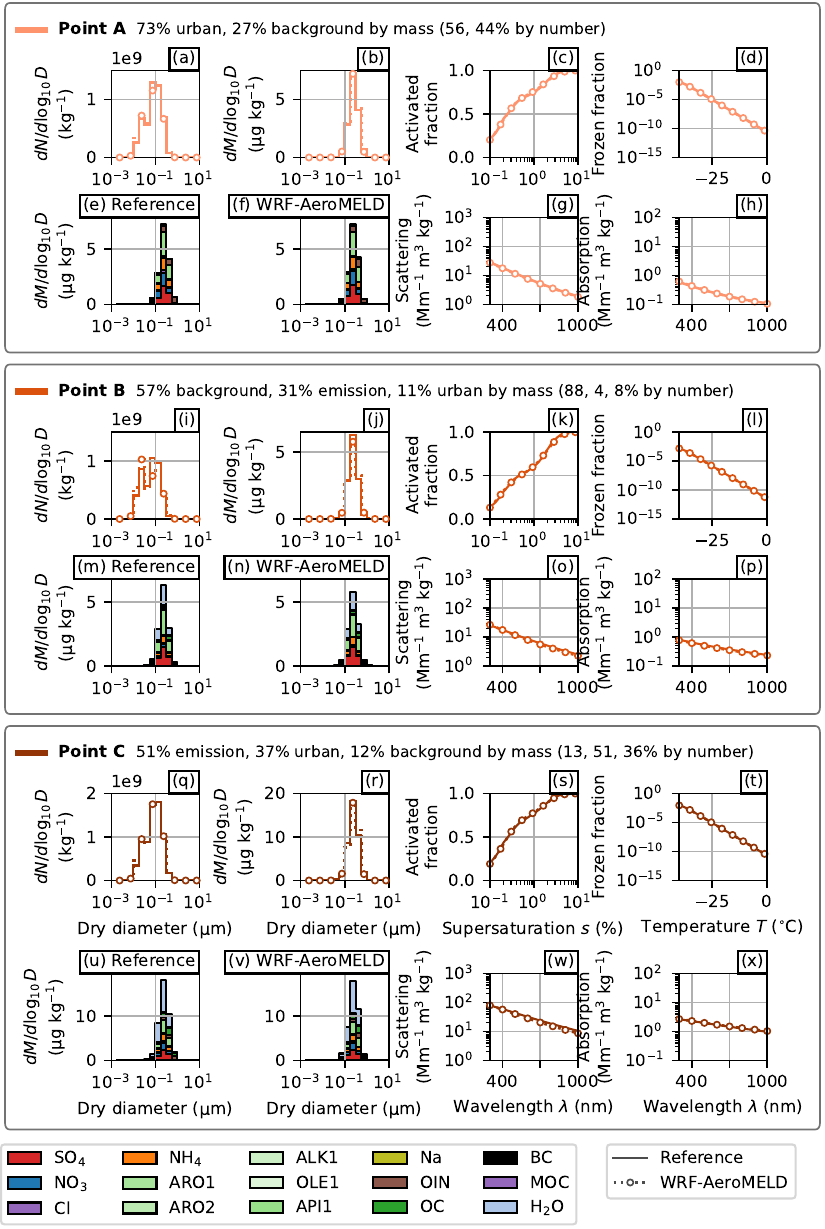}
\caption{WRF-AeroMELD (dotted, open circles) against the
  particle-resolved reference (solid) at sampling points A (a--h), B
  (i--p) and C (q--x) of Figure~\ref{fig:setup} after 12~h; each
  header gives the population mix by mass and, in parentheses, by
  number. To compare the two at the same particle number
  mixing ratio, we multiply the reference number and mass size
  distributions and optical coefficients by the ratio of the
  WRF-AeroMELD to the reference number mixing ratio at each point
  (0.990--1.021).}
\label{fig:point-panels}
\end{figure}

Three sampling points, A, B and~C (Figure~\ref{fig:setup}), represent mixtures
in which a different source population contributes most of the aerosol mass.
The headers of Figure~\ref{fig:point-panels} give their mixtures, and
\ref{app:transport} describes how the points were chosen. The populations differ
in mass per particle. The particles of the urban population have 2.2 times
the mass of those of the background, and those of the emission population
11.8 times, three quarters of it water, so the composition of a grid cell by
mass differs from its composition by number. Figure~\ref{fig:point-panels} compares the number size
distribution, the total and speciated mass size distributions, the CCN
spectrum, the frozen fraction and the optics at each point. The speciated mass
distributions are the most demanding comparison, because they test whether each
species is placed at the right sizes, which determines the hygroscopicity of
the population and its CCN activity. The overall errors at the three points are
2.3--4.1\%, nearly all from the decoder, and the transport errors are
0.2--0.8\%. At~A the largest error is in the speciated mass distribution
(6\%), from decoding both the source populations and their mixture. At~B it is
in the number distribution (10\%), from decoding the background population,
which contributes 88\% of the number there. At~C, where
the emission population contributes most of the mass, the error of every
diagnostic is below 7\%.

\begin{figure}
\noindent\includegraphics[width=\textwidth]{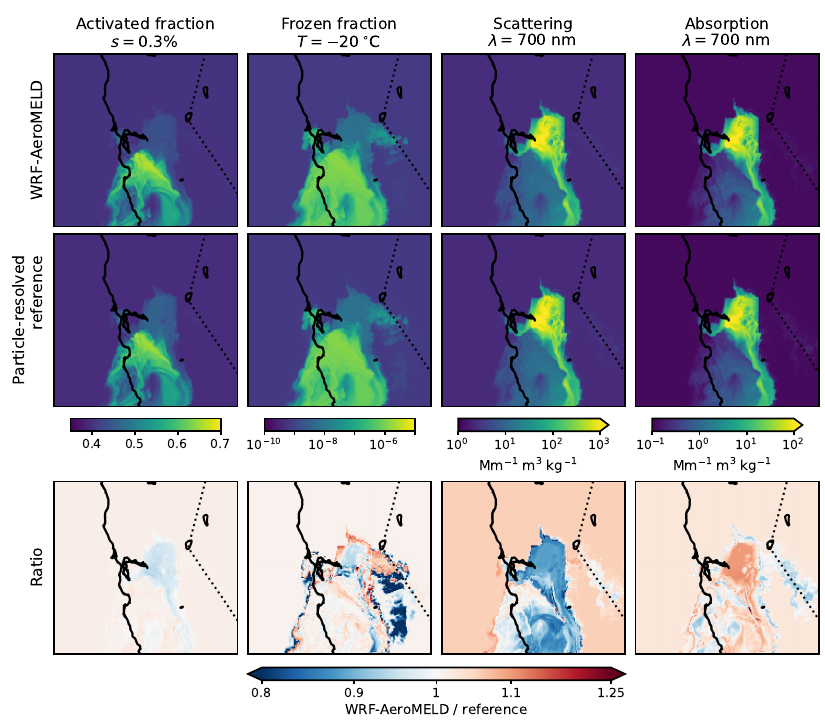}
\caption{Four diagnostics on the lowest model level after 12~h (columns): the
activated fraction at $s=0.3\%$, the frozen fraction at $-20\,^{\circ}$C, and
the scattering and absorption coefficients at 700~nm, per kilogram of dry
air. WRF-AeroMELD (top row), the particle-resolved reference (middle row), and
the ratio WRF-AeroMELD/reference on one color scale (bottom row).}
\label{fig:maps}
\end{figure}

Figure~\ref{fig:maps} maps four of the diagnostics over the domain. Over the
mixed cells the median ratio of the value from WRF-AeroMELD to that of the
reference is 0.91--1.03 for the four diagnostics, and 90\% of the cells lie
between 0.85 and 1.10. Outside the plume the maps show the error of the decoder
for the background population alone, which is uniform in space. There the
scattering is about 5\% too high, and the other three diagnostics are within
3\% of the reference.

\FloatBarrier

\section{Discussion}
\label{sec:discussion}

The sectional and modal schemes lose most of their accuracy in this
test case because they assume that all particles in a bin or mode have
the same composition. Even when the source populations are mixed in
the reference proportions, the sectional and modal schemes have errors
three to five times as large as WRF-AeroMELD's. On the test set, this
difference grows as more populations are mixed. The frozen fraction
shows why the mixing state matters. The dust that nucleates ice is
found on only a few percent of the urban particles, and distributing
it over all particles in a bin or mode increases the frozen fraction
two- to threefold. Adding bins reduces this error only slowly. MATRIX,
whose populations are defined by composition, keeps most of the
dust-containing particles separate, and its frozen-fraction error is a
quarter to a fifth of those of the other three schemes. However,
nearly every particle in this test case contains some black carbon and
organic matter, so MATRIX assigns most particles to a single
population with one composition and a fixed width. The latent state of
WRF-AeroMELD is not tied to size bins or composition classes. With ten
tracers, WRF-AeroMELD reproduces the low frozen fractions that result
when only a few percent of the urban particles carry dust. With many
more tracers, the sectional schemes resolve other features
better. TOMAS-40, with 36 times as many tracers as WRF-AeroMELD,
reproduces the number size distribution as accurately and the
speciated mass size distribution more accurately.

The comparison is designed to test the representations rather than
particular implementations of the schemes. Each sectional or modal
scheme uses its own bins or modes, its own species, its own rules for
assigning particles, and the adjustments that its host models apply
after transport. Adjustments that belong to aerosol processes are
omitted, since no process acts in this test case. The diagnostics of
every scheme are computed from representative particles with the same
particle-level calculations as the reference, and every scheme is
advected in the same way as WRF-AeroMELD. Some of these choices favor
the sectional and modal schemes. Water is advected as a passive
tracer, so these schemes do not need to diagnose it from the relative
humidity, and the test case has no aerosol processes, whose treatment
in these schemes would introduce additional errors. Other choices may
disfavor them. The source populations come from a particle-resolved
library rather than from the emission inventories for which the
schemes were designed, and the fresh carbon of the emission population
is larger than MAM4 expects, although projecting it differently
changes MAM4's error by only about half a percentage point. Finally,
the size distributions are compared in dry diameter. MOSAIC defines
its bins by dry diameter, whereas TOMAS defines them by dry mass per
particle. In wet diameter, which determines the optical properties and
the activation of the particles in a host model, WRF-AeroMELD also
reproduces the size distributions more accurately than every sectional
and modal scheme. A scheme that diagnoses water from the composition
of a bin assigns the same water fraction to every particle in the bin,
so particles of the same dry size but different composition receive
the same wet diameter (\ref{app:baselines}).

Exact mixing in latent space is necessary to transport a learned aerosol state,
but it is not sufficient, and the appendices test two further requirements.
First, the tracers must be advected as accurately as the host model advects any
other tracer. The two alternative transport schemes that we tested keep every
grid cell an exact mixture of its sources, but they do so by advecting the
latent vector less accurately than the number. The resulting smearing of the
plume fronts costs more than exact mixing gains, and the simpler of the two,
donor-cell transport, more than triples the error (\ref{app:schemes}). WRF's
standard tracer advection, in which each tracer has its own flux limiter, moves
the latent vectors slightly away from exact mixtures. This adds less than one
percentage point to the error, does not accumulate over the 12~h, and requires no
change to the host model other than advecting $N\zlat$ in place of $\zlat$.
Second, the decoder must be trained on mixtures, since transport provides it
with the averaged latent vectors of mixed air. A decoder trained only on single
populations more than doubles the error (\ref{app:decoders}).
With WRF's tracer advection, nearly all of WRF-AeroMELD's remaining error comes
from the decoder, through its errors on the single source populations and on
their mixtures. Its largest error is in the speciated mass size distribution,
which requires the decoder to place every species at the right sizes. This is
also the diagnostic in which TOMAS-40, which bins by dry size, has a lower
error.

We chose a passive test case because it allows us to construct a
particle-resolved reference in every grid cell. As long as particle
properties remain fixed, the number mixing ratio contributed by each
source population is sufficient to calculate the diagnostics of the
local mixture. Per-source transport could not be used as a full
aerosol model. It requires the sources to be known, its cost grows
with their number, and it cannot represent an aerosol that ages by
coagulation or condensation. The latent state, in contrast, describes
any population with a fixed number of variables, and aerosol processes
can act on it directly \cite{TangAeroMELDCoag2026}.

This study has several limitations. It covers emission and transport only, in a
single 12-h test case with three source populations, which are held out of
training but come from the same library as the training data. All results are
for the lowest model level. Each source population has 1000 particles. This is
enough for the comparison, because the reference and every representation are
computed from the same particles. However, estimating mixing-state metrics such
as the mixing-state index accurately requires more particles
\cite{Gasparik2020}. The encoder, the sectional and modal schemes and the
reference apply equally to populations with more particles. The sectional and
modal schemes are evaluated as representations, with their adjustments after
transport but without their process code. Future work will add aerosol processes to WRF-AeroMELD as
operators on the latent state, operator-split from transport, beginning with
coagulation \cite{TangAeroMELDCoag2026} and continuing with gas--particle
partitioning and deposition. With these processes included, WRF-AeroMELD and
the sectional and modal schemes can be compared with particle-resolved
simulations in WRF \cite{Curtis2019,Curtis2024}. Because the argument for
advecting the latent state applies to any host model with flux-form transport,
the approach can also be tested in global models such as MPAS
\cite{Skamarock2012,Huang2022}.

\section{Conclusions}
\label{sec:conclusions}

WRF-AeroMELD transports the aerosol number mixing ratio $N$ and the
nine components of $N\zlat$ as ten tracers using WRF's standard
monotonic advection scheme. Dividing the nine transported components
by the local number mixing ratio recovers the latent vector $\zlat$ used by the decoder. We evaluated WRF-AeroMELD in an idealized
12-h simulation over central California with three prescribed particle
populations. Their properties remained fixed, allowing us to calculate
reference diagnostics directly from the source particles and their
transported number mixing ratios. We compared it with the sectional
schemes MOSAIC and TOMAS and the modal schemes MAM4 and MATRIX under
the same transport. Our main conclusions are as follows.
\begin{enumerate}
\item WRF-AeroMELD reproduces the six diagnostics with a mean error of 3.8\%
  after 12~h, compared with 12--19\% for MOSAIC~8, TOMAS-40, MAM4 and MATRIX,
  which advect 25--362 tracers per grid cell compared with WRF-AeroMELD's ten.
\item Advection adds less than one percentage point to the error of every
  representation. The
  differences between the representations therefore arise from how they
  represent the aerosol, and in particular from the assumption that all
  particles in a bin or mode have the same composition.
\item The differences are largest for immersion freezing, which in
  this test case is determined by dust on 4\% of the urban
  particles. MOSAIC, TOMAS and MAM4 spread this dust over all
  particles in a bin or mode and overestimate the frozen fraction two-
  to threefold, and adding bins reduces this error only slowly. MATRIX
  confines the urban dust to its mixed population and has the smallest
  frozen-fraction error of the four comparison schemes, but the error
  is still more than five times that of WRF-AeroMELD.
\item Without the frozen fraction, the errors of the sectional and modal
  schemes are 1.5--3.1 times that of WRF-AeroMELD. TOMAS-40, with 36 times as
  many tracers, reproduces the number size distribution as accurately and the
  speciated mass size distribution more accurately, but all
  four schemes have larger errors than WRF-AeroMELD in the CCN spectrum and the
  optical properties. On mixtures of up to 16 test populations the difference
  increases.
\item Within each family of schemes, more tracers give a lower error, but the
  sectional schemes need more tracers than the modal schemes to reach the same
  error, and WRF-AeroMELD has a lower error than either with fewer tracers.
\item Both WRF's standard tracer advection and a decoder trained on mixtures are
  needed. A transport scheme that keeps mixing exact but advects the latent
  state at first order more than triples the error, and a decoder trained
  without mixtures more than doubles it.
\end{enumerate}
Before WRF-AeroMELD can be used in applications, aerosol processes must be
represented as operators on the latent state, and the approach must be tested in
longer, chemically active and global simulations.

\appendix

\section{Computing the Errors}
\label{app:attribution}

If transport moved every source population exactly, as per-source transport
does, the latent vector of every cell would be the mixture of the sources'
latent vectors $\zlat_s$ in the cell's source fractions,
\begin{equation}
  \zlat=\sum_s f_s\,\zlat_s,
  \label{eq:decomp}
\end{equation}
with number fractions $f_s$ that sum to one, as Equation~\eqref{eq:mix}
gives for exact mixing. Every latent vector would then lie in the triangle
with corners $\zlat_s$, and its position in the triangle would give the
source fractions. An advected latent vector can depart from this ideal in two
ways. It can lie at the wrong position within the triangle, so that the grid
cell contains the source populations, and the number, in the wrong
proportions. It can also leave the triangle and take a value that no mixture of
the sources has. The source fractions implied by any latent vector are its
barycentric coordinates in the plane of the source latent vectors, computed by
least squares.

Each of the transport and representation errors of
Equation~\eqref{eq:budget} can be split into two parts by changing one
ingredient at a time, going from the advected state to the reference through
three intermediate stages. The first is the \emph{clipped mixture}
$\bar\zlat$, the mixture of the sources with the fractions implied by $\zlat$,
set to zero where negative and rescaled to sum to one. The second is the exact
mixture $\zlat^{\mathrm{PS}}$ with the number $N^{\mathrm{PS}}$. The third is
$\mathbf y^{\mathrm D}$, the reference of Equation~\eqref{eq:exact} with the
decoded sources $\mathbf q(\zlat_s)$ in place of $\mathbf q^{\mathrm{PR}}_s$.
Then
\begin{equation}
\begin{split}
  \underbrace{e\big(\Dec(N,\zlat),\,\mathbf y^{\mathrm{PR}}\big)}_{\text{error}}
  \le{}&\underbrace{\underbrace{e\big(\Dec(N,\zlat),\,\Dec(N,\bar\zlat)\big)}_{\text{off-mixture error}}
  +\underbrace{e\big(\Dec(N,\bar\zlat),\,\Dec(N^{\mathrm{PS}},\zlat^{\mathrm{PS}})\big)}_{\text{fraction error}}}_{\text{transport error}}\\[1ex]
  &+\underbrace{\underbrace{e\big(\Dec(N^{\mathrm{PS}},\zlat^{\mathrm{PS}}),\,\mathbf y^{\mathrm D}\big)}_{\text{averaging error}}
  +\underbrace{e\big(\mathbf y^{\mathrm D},\,\mathbf y^{\mathrm{PR}}\big)}_{\text{reconstruction error}}}_{\text{representation error}}.
\end{split}
\label{eq:terms}
\end{equation}
The four terms are as follows.
\begin{itemize}
\item The \emph{off-mixture error} compares the advected latent vector with the
  clipped mixture, both decoded. It is zero for donor-cell transport and
  nearly zero for the shared limiter (\ref{app:schemes}), whose latent vectors
  are mixtures of the sources with fractions very nearly between zero and one.
\item The \emph{fraction error} compares the clipped mixture with the exact
  mixture, each decoded with its own number. It measures the error of having
  the source populations, and the number, in the wrong proportions.
\item The \emph{averaging error} compares the decoded exact mixture with the
  decoded sources mixed in the same proportions. It is zero for an affine
  decoder (\ref{app:decoders}).
\item The \emph{reconstruction error} compares the decoded sources with the
  particle-resolved sources, mixed in the same proportions.
\end{itemize}
By the triangle inequality each error is at most the sum of its two parts, in
every cell and therefore also on average; the terms do not add exactly. The
representation error's two terms use the per-source fractions, so they are
the same for every transport scheme. For per-source transport both transport
terms vanish, and its error is the representation error. The same two
representation terms are measured on mixtures of test populations, with the
test populations in place of the sources: the averaging error compares the
decoded mixture with the mixture of the decoded populations, and the
reconstruction error compares that with the true mixture. For the sectional
and modal schemes, their tracers take the place of $(N,\zlat)$ and their
sources' bins or modes that of $\zlat_s$ (\ref{app:baselines}).

Each stage
of Equation~\eqref{eq:terms} is a set of per-particle diagnostics in the units
of the decoder's output (15 species in 15 size bins, the 15-bin number
distribution, 100 supersaturations, eight wavelengths for each of the optics,
and 100 temperatures), multiplied by the cell's number concentration for the
extensive diagnostics; the mixtures of the sources, $\mathbf y^{\mathrm D}$
and $\mathbf y^{\mathrm{PR}}$, average the sources' per-particle values in the
number fractions of per-source transport. The number concentration per cubic
meter is the number mixing ratio divided by WRF's specific volume of dry air
at the lowest level. The errors are then computed in each mixed cell as on
the test set, at the same supersaturations, temperatures and wavelengths
(Section~\ref{sec:theory-errors}), and averaged over the mixed cells.

\section{Sensitivity to the Transport Scheme}
\label{app:schemes}

To determine how the transport scheme affects the results, we compare WRF's
standard tracer advection, which WRF-AeroMELD uses, with two alternative
schemes that both keep mixing exact.

\paragraph{Exact mixing and flux limiters.} A host model advects $N$ and
$\mlat=N\zlat$ in flux form. On the grid, each time step updates the content
of a cell~$i$ by the fluxes through its faces~$e$,
\begin{equation}
  (\mu c)^{n+1}_i=(\mu c)^n_i-\Delta t\sum_{e\in\partial i}F_e(c),
  \label{eq:fv}
\end{equation}
for every transported quantity $c$, with $\mu$ the dry-air mass. Whether
transport preserves the mixing of Equation~\eqref{eq:mix} depends on how the
face fluxes $F_e$ are computed. If every quantity is moved by the same linear
operator, transport is exact in latent space. Write the number of each source
in a cell as $N_s=Nf_s$, so that $N=\sum_s N_s$ and
$\mlat=\sum_s N_s\zlat_s$. Applying a linear operator $\mathcal L$ to $N$ and
to every component of $\mlat$ gives
\begin{equation}
  \zlat^{n+1}=\frac{\mathcal L(\mlat^n)}{\mathcal L(N^n)}
  =\sum_s\frac{\mathcal L(N^n_s)}{\mathcal L(N^n)}\,\zlat_s,
  \label{eq:linear-mix}
\end{equation}
which is Equation~\eqref{eq:decomp} with the fractions that transporting each
source by $\mathcal L$ would give: the latent state stays exactly a mixture of
the sources, with fractions that sum to one. High-order linear schemes,
however, overshoot at sharp fronts, which would push the fractions below zero
or above one, and the only linear schemes that never overshoot are first
order (Godunov's theorem; \citeA{durran2010numerical}). Host models therefore
limit their fluxes. WRF's monotonic option, used for chemical tracers
\cite{Skamarock2006,Wang2009}, is a flux-corrected transport scheme
\cite{Zalesak1979}: each face flux is a first-order upwind flux plus a
fraction $\alpha_e\in[0,1]$ of the antidiffusive difference between a
high-order flux and the upwind one,
\begin{equation}
  F_e=F^{\mathrm{up}}_e+\alpha_e\big(F^{\mathrm{high}}_e-F^{\mathrm{up}}_e\big),
  \label{eq:fct}
\end{equation}
with $\alpha_e$ given by Zalesak's procedure, which keeps every cell within the
range of its own and its upwind neighbors' old values. The limiter makes the
operator nonlinear, and the three transport schemes differ in how $N$ and
$\mlat$ share it.

\paragraph{Three transport schemes.} The three schemes differ in how the
latent vector is treated by WRF's limiter.
\begin{itemize}
\item \textbf{Transport scheme 1: donor-cell.} $N$ is advected with WRF's
  monotonic scheme, and the latent vector is advected with the fluxes of $N$,
  taking the value of the upwind cell. Every new latent vector is then an
  average of old ones, so every grid cell remains an exact mixture of the
  sources. However, the composition is advected by first-order upwind
  transport, which diffuses fronts.
\item \textbf{Transport scheme 2: shared limiter.} $N$ and the nine components
  of $N\zlat$ use one limiter coefficient per face, chosen so that $N$ keeps
  WRF's bounds and each component of $\zlat$ stays within the range of the
  cell and its upwind neighbors. Because every tracer moves with the same
  coefficients, mixing stays exact, as for donor-cell transport, and the
  bounds keep the latent vector within the triangle of the mixtures, or very
  nearly. However, each face must satisfy the bounds of every tracer at once,
  so the scheme limits the fluxes more often than any single tracer requires.
\item \textbf{Transport scheme 3: separate limiters (WRF-AeroMELD).} $N$ and
  each component of $N\zlat$ are advected as ordinary tracers, each with its
  own limiter coefficients. This is WRF's standard tracer advection, and the
  scheme WRF-AeroMELD uses inside WRF (Section~\ref{sec:theory-transport}). It
  keeps the fronts sharpest, but the different coefficients break exact
  mixing, so the latent vector can leave the triangle.
\end{itemize}
The three schemes trade off the two parts of the transport error
(\ref{app:attribution}). Donor-cell transport avoids off-mixture error at the
cost of a large fraction error. Separate limiters keep the fraction error small
at the cost of some off-mixture error. The shared limiter keeps mixing exact,
with a fraction error smaller than that of donor-cell transport but larger
than that of separate limiters.

\paragraph{Implementation.} \emph{Transport scheme~1, donor-cell:} the latent
update is
$(\mu\mlat)^{n+1}_i=(\mu\mlat)^n_i-\Delta t\sum_e F_e(N)\,\zlat^n_{\mathrm{up}(e)}$,
with $F_e(N)$ the number's limited face flux and $\mathrm{up}(e)$ the cell the
flux leaves. Since $N$ is updated with the same fluxes, each new latent vector
is a convex combination of the old ones in the cell and its upwind neighbors,
so every cell stays a mixture of the sources with fractions between zero and
one. \emph{Transport scheme~2, shared limiter:} a single coefficient per face
that satisfies, in both cells it separates, the bounds on $N$ and the linear
constraints $z_{\min,c}N\le(\mlat)_c\le z_{\max,c}N$ for every component $c$,
computed with Zalesak's procedure for each constraint and the minimum over
constraints. Given the coefficients the operator is linear, so
Equation~\eqref{eq:linear-mix} holds, and the bounds keep the fractions close
to the range from zero to one. \emph{Transport scheme~3, separate limiters:}
the monotonic scheme applied to $N$ and to each component of $\mlat$, with
independent coefficients, which break Equation~\eqref{eq:linear-mix}.
\emph{Per-source transport:} the monotonic scheme applied to each source's
number separately, in five classes (initial background, initial urban,
background inflow, and two emission inputs, both of which carry the emission
population).

\begin{figure}
\noindent\includegraphics[width=\textwidth]{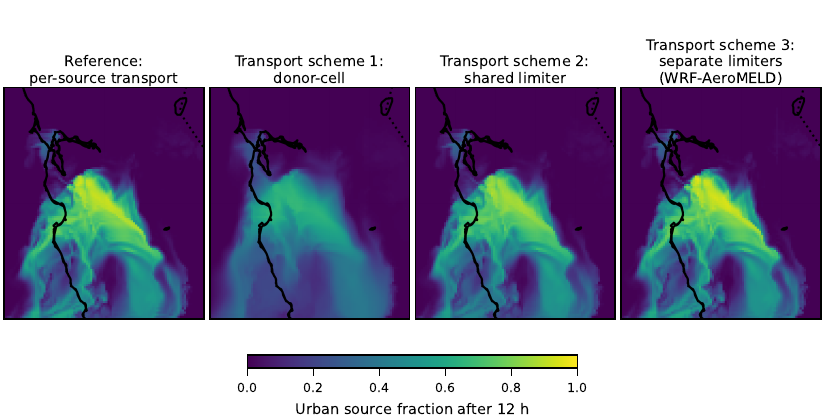}
\caption{Urban source fraction (by number) on the lowest model
level after 12~h, around the urban plume, for per-source transport (the
reference) and for the latent state of decoder~3 (representative seed) under transport schemes~1
(donor-cell), 2 (shared limiter) and~3 (separate limiters) (the fractions
implied by the latent vector,
Equation~\eqref{eq:decomp}).}
\label{fig:fronts}
\end{figure}

\paragraph{Fronts and the two transport errors.} The transport schemes differ
most at the plume fronts, where cells mix (Figure~\ref{fig:fronts}).
Donor-cell transport smears the fronts: the mean gradient of the urban
fraction over the mixed cells is less than half that of per-source transport,
and the filaments inside the plume are lost. The shared limiter retains most
of these filaments, and separate limiters retain all of them. Even separate limiters differ from
per-source transport in the number, since they advect the number as one tracer
and per-source transport advects each source's number separately: the limiter
acts on the total differently from its parts, and the two numbers differ by
1.4\% (RMS over the domain) and 2.9\% over the mixed cells, where the fronts
are. This difference is part of the fraction error.

\begin{figure}
\centering
\includegraphics[width=0.6\textwidth]{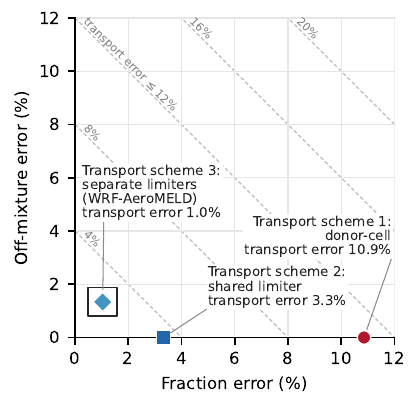}
\caption{The two parts of the transport error, Equation~\eqref{eq:terms}, for
the three transport schemes with decoder~3, after 12~h: the off-mixture error
against the fraction error, overall errors over the mixed cells, medians over
ten training seeds. Each label gives the scheme's transport error measured
directly. Along each gray diagonal the two parts have a constant sum, which
bounds the transport error, as marked on the diagonal from the upper left
corner. WRF-AeroMELD's scheme, transport scheme~3, is boxed. The two axes are
the same.}
\label{fig:tradeoffs-transport}
\end{figure}

Figure~\ref{fig:tradeoffs-transport} shows the two parts of the transport
error of each scheme with decoder~3, which is trained on mixtures of up to 16
populations. Donor-cell transport keeps every cell an exact mixture but
smears the fronts, so its transport error is all fraction error, and large.
The shared limiter also keeps mixing exact, with a smaller fraction error.
Separate limiters move the source fractions most accurately, to within 0.01
of per-source transport (RMS), but they depart from the exact mixtures by 1.3\% of the
spacing of the source latents (RMS; 5.4\% at the 99th percentile, with
implied fractions as low as $-0.05$). This small departure costs less than
the shared limiter's extra diffusion, and separate limiters have the smallest
transport error. With decoder~3, the error is 12.2\% with donor-cell
transport, 5.4\% with the shared limiter and 3.9\% with separate limiters
(medians over ten seeds), compared with a representation error of 3.5\%, the
error that remains when transport is exact.

The frozen fraction is the diagnostic most affected by smeared fronts. Its
error is 24\% with donor-cell transport, compared with 1.9\% with separate
limiters. Nearly all
ice-active particles are in the urban population, whose frozen fraction at
$-35\,^{\circ}$C is about 2000 times the emission population's and
$2\times10^{4}$ times the background's, so a cell's frozen fraction is about
$f_{\mathrm{urban}}$ times the urban value, and its relative error is the
relative error in $f_{\mathrm{urban}}$, largest at the urban plume's edges.
The optics and the speciated mass distribution follow, and the number size
distribution and the CCN spectrum depend least on the transport scheme.

\begin{figure}
\centering
\includegraphics[width=0.6\textwidth]{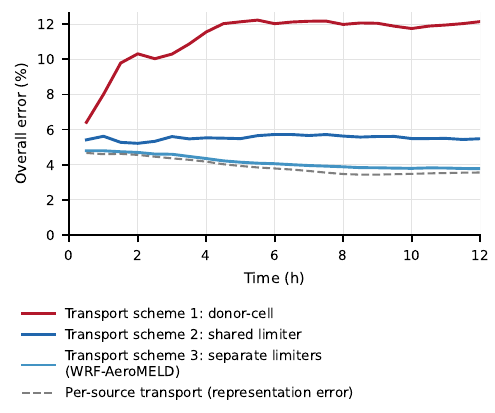}
\caption{Overall error over the mixed cells through the 12-h run of
  each transport scheme with decoder~3 (representative seed). The
  dashed line is the representation error, the error of per-source
  transport, which defines the reference mixing proportions and
  therefore has zero transport error by construction.}
\label{fig:time}
\end{figure}

\begin{figure}
\noindent\includegraphics[width=\textwidth]{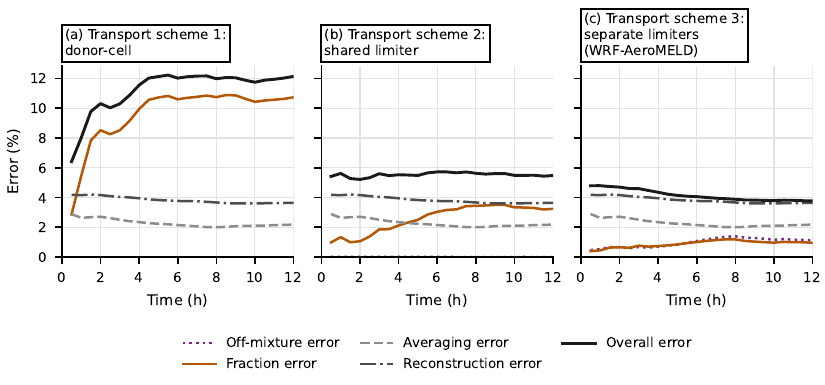}
\caption{The four terms of Equation~\eqref{eq:terms} and the error through the
12-h run for (a)~transport scheme~1, donor-cell, (b)~transport scheme~2, the
shared limiter, and (c)~transport scheme~3, separate limiters, each with
decoder~3 (representative seed); overall errors over the mixed surface cells.
The averaging and reconstruction errors (the decoder's) are the same in every
panel.}
\label{fig:terms}
\end{figure}

\paragraph{Through the run.} Figure~\ref{fig:time} follows each transport scheme's error
through the 12-h run, as the number of mixed cells grows from about 1400 after
30~min to about 6000 after 8~h. Donor-cell transport's error grows as the
fronts thicken over the first 4~h, and then stays there. With separate
limiters the error stays within 0.5 percentage points of per-source
transport's throughout, and both transport terms stay below 2\%
(Figure~\ref{fig:terms}); the off-mixture error rises over the first 8~h, as
the mixed cells spread, and then falls: over the 12~h of this limited-area
test case the departure from exact mixing does not accumulate. The shared limiter lies
between. With the affine decoder, separate limiters stay within about 0.2
percentage points of its representation error throughout.

\begin{figure}
\noindent\includegraphics[width=\textwidth]{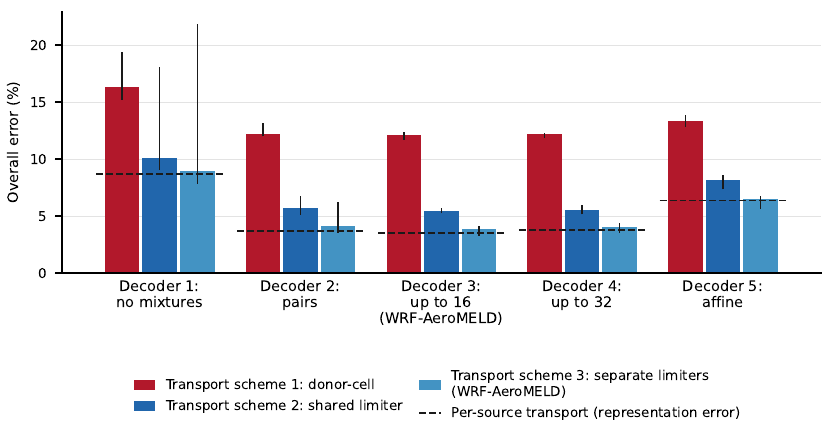}
\caption{Overall error after 12~h over the mixed cells of the five decoders
(groups of bars) under the three transport schemes (colors): median over ten
training seeds (bars) with the range (whiskers). The dashed line in each group
is the decoder's representation error, the error of per-source transport
(median over seeds).}
\label{fig:models}
\end{figure}

\paragraph{All five decoders.} Figure~\ref{fig:models} repeats the comparison
for the five decoders of \ref{app:decoders}. Under donor-cell transport the
transport error dominates, and decoders~2, 3 and~4 have nearly the same error.
With separate limiters the median error of every decoder is within 0.5
percentage points of its representation error, and the differences between
the decoders on the test set reappear. The shared limiter adds 1.4 to 2.1
percentage points to the median representation error of every decoder.

\section{Sensitivity to the Decoder}
\label{app:decoders}

WRF-AeroMELD's decoder is trained on mixtures of up to 16 populations. To
determine how the decoder affects the results, we compare it with four
alternatives, named by how they were trained (\ref{app:training}):
\begin{itemize}
\item \textbf{Decoder 1: no mixtures.} A nonlinear decoder, a multilayer
  perceptron (MLP), trained on single populations only.
\item \textbf{Decoder 2: pairs.} The same, trained also on mixtures of two
  populations.
\item \textbf{Decoder 3: up to 16 (WRF-AeroMELD).} The same, trained on
  mixtures of up to 16 populations.
\item \textbf{Decoder 4: up to 32.} The same, trained on mixtures of up to
  32 populations.
\item \textbf{Decoder 5: affine.} An affine map of the latent vector,
  trained also on mixtures of two populations.
\end{itemize}
Training on mixtures asks a decoder to reproduce the diagnostics of mixtures
of training populations from their averaged latent vectors
\cite{zhang2018mixup,SalehAeroMELD2026}, which lowers its averaging error. The
affine decoder,
\begin{equation}
  \mathbf q(\zlat)=A\,\zlat+\mathbf b,
  \label{eq:affine}
\end{equation}
in physical units, with $A$ and $\mathbf b$ nonnegative, has no averaging error
at all, because an affine map of an average is the average of the maps
(Equation~\eqref{eq:decoder-commute}); but it is less flexible than a nonlinear decoder in
reconstructing single populations. The five decoders thus trade averaging error
against reconstruction error.

\paragraph{Decoding mixtures.}
For any per-particle diagnostic $q$, a fraction of particles or a per-kilogram
diagnostic divided by $N$, the true value for a mixture of the source
populations with number fractions $f_s$ is the average of the sources'
values,
\begin{equation}
  q=\sum_s f_s\,q_s.
  \label{eq:diag-mix}
\end{equation}
The latent vector of the mixture averages in the same way,
Equation~\eqref{eq:decomp}, but a decoder need not map the one average to the
other. The affine decoder of Equation~\eqref{eq:affine} does, for every
mixture, because the fractions sum to one:
\begin{equation}
  \mathbf q\Big(\sum_s f_s\zlat_s\Big)
  =A\sum_s f_s\zlat_s+\mathbf b
  =\sum_s f_s\,\mathbf q(\zlat_s),
  \label{eq:decoder-commute}
\end{equation}
so its averaging error, Equation~\eqref{eq:terms}, is zero. A nonlinear decoder
departs from Equation~\eqref{eq:decoder-commute} by an amount set by its
curvature between the mixed latent vectors. Training on mixtures reduces this
departure on mixtures like those it is trained on; the averaging error
measures what remains.

\begin{figure}
\centering
\includegraphics[width=0.6\textwidth]{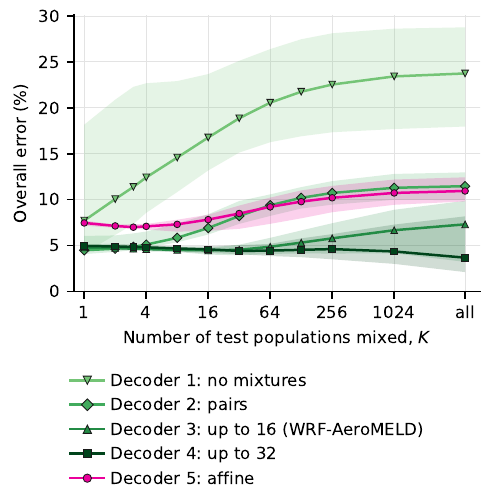}
\caption{Error of each decoder on mixtures of $K$ test populations with
Dirichlet weights, from single populations ($K=1$) to the whole test set
($K=5000$): the overall error (mean over the six diagnostics) against the
true mixture, mean over ten seeds (lines) with the range over seeds
(bands).}
\label{fig:kcurves}
\end{figure}

\paragraph{The test set and the test case.} Figure~\ref{fig:kcurves} shows how
well each decoder reproduces mixtures of test populations. On single test
populations ($K=1$), which measure the reconstruction error alone, the
nonlinear decoders trained on mixtures do best; the affine decoder, less
flexible, has about one and a half times their error, and decoder~1's mean is raised by three
of its ten seeds, which reconstruct poorly, and to a lesser degree by a fourth. As populations are mixed, the
decoders separate by their averaging error. Decoder~1, trained without
mixtures, has an error that grows with~$K$, almost all of it averaging error.
This is the effect of the mixtures, not of the Kullback--Leibler term that
this decoder also lacks: decoder~3 trained without the term has errors close
to decoder~3's, with overlapping ranges over seeds at every~$K$. Decoder~2, trained on pairs, has
an error that grows more slowly, and is within 0.3 percentage points of
decoder~3's for mixtures of up to three populations, as many as the test case
mixes. Decoders~3 and~4 stay nearly level up to the 16 and 32 populations they
are trained on. Beyond that, decoder~3's error grows, to 7.3\% for the whole
test set, almost all of it in the speciated mass distribution, while
decoder~4's falls to 3.7\%. The affine decoder decodes every
mixture exactly as the mixture of its decodes, so its error is that of its
single populations, mixed. It too rises for large~$K$ in the
speciated mass distribution: in mixtures of many populations the mass is
dominated by the few populations with the most mass, so their errors no
longer average out.

\begin{figure}
\centering
\includegraphics[width=0.6\textwidth]{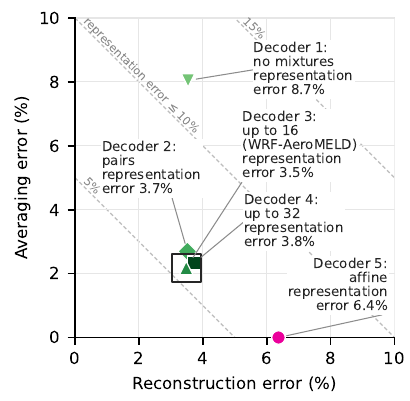}
\caption{The two parts of the representation error, Equation~\eqref{eq:terms},
for the five decoders in the test case after 12~h, on the exact mixtures of
per-source transport: the averaging error against the reconstruction error,
overall errors over the mixed cells, medians over ten training seeds;
decoders~2--4 lie close together. Each label gives the decoder's
representation error measured directly. Along each gray diagonal the two
parts have a constant sum, which bounds the representation error, as marked
on the diagonal from the upper left corner. WRF-AeroMELD's decoder,
decoder~3, is boxed. The two axes are the same.}
\label{fig:tradeoffs-decoders}
\end{figure}

Figure~\ref{fig:tradeoffs-decoders} shows the two parts of the representation
error in the test case, on the exact mixtures of per-source transport after 12~h.
The ranking is that of the test set: training on mixtures lowers the averaging
error, which the affine decoder does not have at all, while the nonlinear
decoders' reconstruction errors are about half the affine decoder's.
Decoders~2, 3 and~4 have nearly the same representation error in the test
case (medians of 3.7, 3.5 and 3.8\%), which mixes at most three sources.
WRF-AeroMELD uses decoder~3 because it stays as accurate on mixtures of up to
16 populations, where decoder~2's error grows (Figure~\ref{fig:kcurves}); the
affine decoder trades some accuracy for a guarantee that mixtures are decoded
exactly. Under every transport scheme decoder~3 has the lowest error and
decoders~5 and~1 the highest (Figure~\ref{fig:models}).

\section{The WRF Simulation and the Offline Transport}
\label{app:transport}

\paragraph{WRF's scalar advection.} WRF advances scalars with a three-stage
Runge--Kutta scheme \cite{Wicker2002,Skamarock2008}. The first two stages use
unlimited fluxes of fifth order in the horizontal and third order in the
vertical, reduced to third and second order next to the lateral boundaries and
to second order next to the bottom and top; the final stage applies the
monotonic limiter of Equation~\eqref{eq:fct} \cite{Skamarock2006,Wang2009}. Each
quantity $c$ is updated as $(\mu c)^{n+1}=(\mu c)^n+\Delta t\,m_y T(F)$, with
$T(F)$ the tendency (minus the flux divergence) including the map factors $m_x$, $m_y$; the
limiter's bounds for a cell are its own old value and those of its upwind
neighbors, and each face's coefficient is the smaller of the inflow factor of
the receiving cell and the outflow factor of the donor cell \cite{Zalesak1979}.
The lateral boundary cells take the boundary value at inflow and the
neighboring interior value at outflow.

\paragraph{The WRF simulation.} The domain has $169\times159$ cells at 4~km
($676\times636$~km) and 39 levels, on a Lambert conformal projection centered
at $38^{\circ}$N, $121.5^{\circ}$W. WRF version~4.7.1 \cite{Skamarock2019}
is run with a 20~s time step, the 39 vertical levels of an earlier WRF~3.9 configuration of the test case, a terrain-following
dry-air mass coordinate, no boundary-layer scheme and no explicit diffusion.
The meteorology starts at 00~UTC from 6-hourly North American Mesoscale (NAM)
analyses on 39 pressure levels up to 50~hPa, which also give the lateral
boundary conditions, and the model top is at 50~hPa. The physics are the
two-moment microphysics of \citeA{Morrison2009}, the RRTMG longwave and
shortwave radiation \cite{Iacono2008} every 30~min, the Kain--Fritsch cumulus
scheme \cite{Kain2004}, the Eta similarity surface layer \cite{Janjic1994}
and the five-layer thermal diffusion land surface model. These physics shape
only the meteorology: the aerosol is passive, with no feedback on radiation or
clouds, and its transport uses only WRF's air-mass fluxes.

\paragraph{The offline transport.} WRF was rerun writing its time-averaged
air-mass fluxes every minute; the rerun's dry-air mass is identical to that of
the runs it replaces to the last bit. Every representation is transported on
these fluxes, with the same initial and boundary conditions and emissions,
the tracers of each computed from the same three source populations. The offline transport
uses WRF's 20~s time step, with the fluxes of each minute for its three steps
and all three Runge--Kutta stages, and advances the air mass with the same
fluxes. Emissions are added after each step at the lowest level as the number
flux times $\alpha\,\Delta t/\Delta z$, with $\alpha$ the specific volume of dry
air, as in WRF-Chem. A 12-h run of every transport scheme with the source latents
of all five decoders takes about 17 minutes on one GPU, and a run of a
sectional or modal scheme, with up to 362 tracers, up to about 75 minutes. Run inside WRF,
WRF-AeroMELD with separate limiters agrees with the offline transport at every
15-min output of the 12~h: the source fractions implied by the latent
vectors differ by less than $10^{-3}$ (RMS over the mixed cells), the
number by less than $6\times10^{-4}$ in $\log_{10}$, and the offline air mass
matches WRF's to five digits throughout the run. After 12~h the overall error
against the reference is 3.78\% both online and offline. Besides these checks,
the offline transport reproduces an earlier online implementation of a
donor-cell scheme in WRF to better than $5\times10^{-4}$ in the implied source
fractions, when it weights the vertical inflow as that implementation does (by
the donor layer's thickness rather than the receiving layer's); the donor-cell
scheme of this paper uses the consistent weighting, the receiving layer's.

\begin{figure}
\noindent\includegraphics[width=\textwidth]{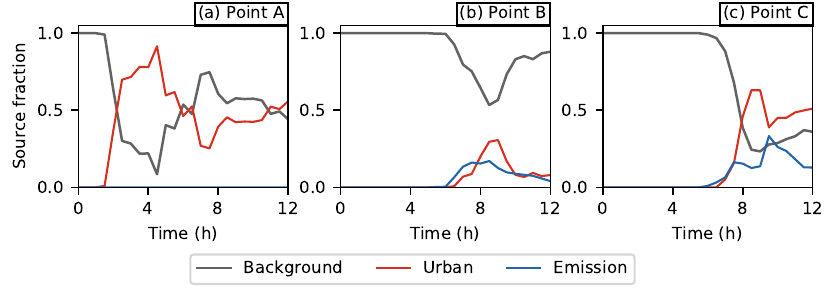}
\caption{Source fractions (by number) on the lowest model level of each
source population (background, urban and emission) through the 12-h run, by
per-source transport, at sampling points (a)~A, (b)~B and~(c)~C of
Figure~\ref{fig:setup}.}
\label{fig:source-fractions}
\end{figure}

\paragraph{The sampling points.} Each of the sampling points A, B and~C is
chosen at 12~UTC in per-source transport among the cells in which one source
population holds 50--75\% of the mass, water included, and another at least 20\%, at least 10
cells from the lateral boundaries, as the one with the smallest $3\times3$
standard deviation of the mass fractions, so that it is not on a sharp front
in space. Figure~\ref{fig:source-fractions} shows how their mixtures developed
through the run.

\section{The Sectional and Modal Schemes}
\label{app:baselines}
This appendix describes how we use MOSAIC, TOMAS, MAM4, and MATRIX in
the transport comparison. For each scheme, we first assign the
particles in each of the three PartMC source populations to bins or
modes and calculate the corresponding particle numbers and species
masses. These quantities define the aerosol supplied through the
initial conditions, boundary inflow, and emissions. Each scheme then
transports its own tracers. To compare its predictions with the
particle-resolved reference, we calculate aerosol diagnostics from the
size distribution and composition represented by the transported bins
or modes.

\paragraph{Assigning source particles to bins and modes.}
For each bin or mode, we sum the number weights of the particles
assigned to it and their corresponding species masses, using the
species groupings of each scheme. This assignment preserves total
particle number and the mass of each species carried by the scheme. We
then divide these sums by the total number of particles represented by
the entire source population.  The resulting state gives, for each bin
or mode, its fraction of the total particle number and its
contribution to the mean mass per particle of each
species. Multiplying this state by the local number mixing ratio $N$
gives the number and mass mixing ratios used for transport. When
populations mix, their normalized states are averaged in proportion to
the number of particles each contributes, as for $\zlat$.

\paragraph{Calculating diagnostics from bins and modes.}
After transport, we calculate aerosol diagnostics using representative
particles of different sizes. Within a bin or mode, all representative
particles have the same species mass fractions, including the water
mass fraction. We treat water as a passive species throughout this
experiment.  For the sectional schemes, we adapt the within-bin
reconstruction used in the AeroMELD coagulation study
\cite{TangAeroMELDCoag2026}. We use 20 representative sizes per bin,
including both edges, spaced uniformly in log dry diameter for MOSAIC
and in log dry mass for TOMAS. We assign maximum-entropy weights that
reproduce the particle number in each bin and either its mean dry
volume per particle (MOSAIC) or its mean dry mass per particle
(TOMAS). For the modal schemes, the particle number, dry volume, and
prescribed width of each mode determine its number-median dry
diameter. We sample the resulting lognormal distribution at four sizes
per diagnostic size bin. In both cases, the weighted representative
particles reproduce the number, dry volume, and species masses of
their bin or mode.

\paragraph{MOSAIC 8.}
We use the eight-bin configuration of the MOSAIC aerosol module as
implemented in WRF-Chem. Its dry-diameter bin edges are 0.039, 0.078,
0.156, 0.3125, 0.625, 1.25, 2.5, 5, and 10~$\mu$m. The 15 species in
our source populations follow MOSAIC's species definitions
\cite{Zaveri2008} and include water and five secondary organic
species.  Each bin carries particle number and all 15 species masses,
giving 128 tracers. WRF-Chem configurations with secondary organic
aerosol carry 10--17 species per bin. We assign particles outside the
size range to the nearest end bin. In particular, 36\% of the
background particles, 13\% of the urban particles, and 26\% of the
emission particles fall below 39~nm and contribute their number and
mass to the first bin.

In WRF-Chem, MOSAIC's section-moving routine transfers a bin's contents
to the appropriate size bin if its mean dry volume per particle falls
outside the bin bounds. It then adjusts particle number as needed to
keep the mean within bounds. Mixing averages each bin's mean in number,
which keeps it inside the bin. Accordingly, at every half-hourly surface output, all
bins containing more than $10^{-12}$ of their grid cell's particles had
means within bounds. We omitted section moving because its estimated
effect was less than $10^{-17}$ of the total particle number.

\paragraph{TOMAS-40.}
We use the 40-bin TOMAS configuration of GEOS-Chem~14.8
\cite{Lee2013,Trivitayanurak2008}. Following the TOMAS bin definition,
we assign each PartMC particle to a bin using the combined mass of its
sulfate, sea salt, black carbon, organic matter, and dust. Successive bin
edges double in mass, from $9.8\times10^{-25}$ to $1.1\times10^{-12}$~kg
(approximately 1~nm to 10~$\mu$m in dry diameter). The three source
populations occupy bins 8--28. Each bin carries particle number, water,
and the masses of the five dry components listed above, with separate
hydrophobic and hydrophilic classes for both black carbon and organic
matter. This gives 360 bin tracers. Bulk ammonium and nitrate bring the
total to 362.

We combine sodium and chloride as sea salt and assign all secondary
organic species to hydrophilic organic matter. GEOS-Chem prescribes fixed
hydrophobic fractions for emitted carbon and ages it on a timescale of
about one day. Here, we assign a PartMC particle's black carbon and
primary organic matter to the hydrophobic classes if it meets MAM4's
fresh-carbon criterion and to the hydrophilic classes otherwise. This
criterion accounts for the amount and composition of the other dry
material on the particle, as described below. In the emission population,
96\% of the black carbon mass and 97\% of the primary organic mass are
classified as hydrophobic. Because a bin has one composition, this split
decides only where the organic carbon of aged particles goes.

After each transport step, we apply TOMAS's correction for bins whose
mean dry mass per particle falls outside the bin bounds. The correction
redistributes particles among bins. Mixing cannot move a bin's mean out
of the bin, but the separate limiters can, in nearly empty bins at the
plume edges. Without the correction, the bins whose means have left their
bounds hold about $10^{-6}$ of the particles at 12~h.
Applying the correction increases the transport error from 0.4\% to
0.7\% and reduces the overall error by 0.2 percentage points, mainly
through a smaller frozen-fraction error. We omit water equilibration,
condensation, nucleation, coagulation, and carbon aging in this passive
test case.

For the diagnostics, we distribute bulk ammonium and nitrate among the
bins in proportion to their sulfate mass. We calculate each representative
particle's dry diameter from the volumes of all its dry species, including
ammonium and nitrate. The wet diameter also includes water. We use the
same optical and hygroscopic properties for both black carbon classes
and treat hydrophilic organic matter as secondary organic aerosol.

\paragraph{MAM4.} We use the four lognormal modes of E3SM's MAM4
\cite{Liu2016}. The accumulation and coarse modes both have a fixed
geometric standard deviation of 1.8 and carry sulfate, primary and secondary
organic matter, black carbon, dust, and sea salt. Their number-median dry
diameters range from 53.5 to 440~nm and from 1 to 4~$\mu$m, respectively.
The Aitken and primary-carbon modes both have a fixed geometric standard
deviation of 1.6. The Aitken mode (8.7--52~nm) carries sulfate, secondary
organic matter, and sea salt. The primary-carbon mode (10--100~nm) carries
primary organic matter and black carbon. Including particle number and
passive water in each mode gives 25 tracers.

We assign each PartMC source particle to a mode. For particles containing
black carbon or primary organic matter, we apply MAM4's aging criterion
to the individual particle. We treat its black carbon and primary organic
matter as the core and weight the volume of each remaining dry constituent
by its hygroscopicity relative to sulfate. If the resulting coating is
equivalent to fewer than eight monolayers of sulfate on the core, we assign
the particle to the primary-carbon mode. We assign all other particles by
dry diameter. Those smaller than 53.5~nm enter the Aitken mode, those at or
above 1~$\mu$m enter the coarse mode, and those in between enter the
accumulation mode.

We combine nitrate and ammonium with sulfate, the five secondary organic
species into secondary organic matter, the two primary organic species into
primary organic matter, and sodium and chloride into sea salt. If the
assigned mode does not carry a species, we place that species' mass in the
accumulation mode. The particle number stays in the assigned mode.

After each 20-s transport step, we apply MAM4's size adjustment, which
adjusts modal number on a one-day timescale and allows particle transfer
between the Aitken and accumulation modes. In the emission population,
fresh carbon particles have a median dry diameter of about 220~nm. Assigning
them to the primary-carbon mode puts its number-median dry diameter above
the 100~nm upper bound, which the carbon emitted in E3SM does not exceed.
The adjustment increases particle number at fixed
dry volume, bringing the median diameter back toward the bound. Applied to
the emission population alone for 12~h, the adjustment doubles its total
particle number.

We tested two changes to the default MAM4 configuration. Assigning fresh carbon
particles larger than 100~nm to the accumulation mode keeps all modal
medians within their bounds. The size adjustment then has no effect, and
the overall error after 12~h is 19.5\%, compared with 19.0\% in the baseline
run. In a separate test, retaining the original mode assignments but carrying
all 15 PartMC species in every mode reduces the overall error to 16.9\%.
The error in the speciated mass distribution falls from 27.6\% to 19.2\%,
and the error in the CCN spectrum falls from 15.8\% to 12.0\%.

\paragraph{MATRIX.} We use MATRIX mechanism~1 \cite{Bauer2008}, as
implemented in GISS ModelE~2.1.2. It represents the aerosol with 16 lognormal
populations distinguished by composition and, in some cases, by size. The
geometric standard deviation is fixed at 1.6 for Aitken sulfate, 2.0 for
coarse sea salt and the mixed population, and 1.8 for the other populations.
Each population carries particle number and the masses of the species it
contains. Sulfate, black carbon, organic matter, dust, and sea salt are
tracked by population. Nitrate, ammonium, and water are carried as bulk
tracers, giving 53 tracers in all. For the diagnostics, we distribute nitrate
and ammonium among populations in proportion to their sulfate mass, and water
in proportion to their combined sulfate and sea salt mass.

Two choices are needed to map PartMC species to MATRIX. We group sodium and
chloride as sea salt up to the NaCl stoichiometric ratio and carry any excess
chloride mass with bulk nitrate. We also combine all primary and secondary
organic species into a single organic matter tracer in each population that
carries organics. For the diagnostics, we give this organic matter the
properties of PartMC's secondary organic aerosol, which most closely matches
the hygroscopicity of MATRIX organic matter. Using the properties of PartMC's
primary organic carbon instead raises the representation error for the
12-h mixtures from 11.8\% to 14.1\%, mainly through the speciated mass
distribution.

We use MATRIX's coagulation table to assign each PartMC particle to a
population according to its composition. We first check for black carbon,
organic matter, dust, and sea salt, counting even trace amounts. Particles
containing none of these enter the Aitken sulfate population if their wet
diameter is below 0.1~$\mu$m and the accumulation sulfate population
otherwise. When only one of the four constituents is present, organic matter
has its own population, sea salt is classified by size, black carbon by
coating, and dust by both size and coating. Two combinations have dedicated
populations: black carbon with organic matter and black carbon with dust.
All other combinations enter the mixed population. With this assignment,
86--96\% of the particles in each source enter the black-carbon--organic
population. All urban dust enters the mixed population, which contains
4.5\% of urban particles by number.

After each transport step, we apply three MATRIX adjustments. These
recalculate sea salt particle number from sea salt mass, transfer whole dust
or black carbon populations to a more heavily coated class when a coating
threshold is exceeded, and transfer particles from Aitken to accumulation
sulfate. The populations on which these adjustments act remain empty
throughout this experiment, so the adjustments have no effect.

For the source mixtures present at 12~h, mechanisms~2, 3, and~5--7 give
the same representation as mechanism~1. Mechanisms~4 and~8 lack a dedicated
black-carbon--organic population. They assign those particles to the mixed
population, spreading the urban dust over a larger fraction of particles.
The overall representation error rises from 11.8\% to 21.1\%, and the
frozen-fraction error rises from 9.7\% to 55.1\%.

\paragraph{Size distributions by wet diameter.}
In the main comparison, we group particles by dry diameter when calculating
the number and speciated mass distributions. Here we use wet diameter and
keep the same 15 diagnostic bin edges. Water contributes to the speciated
mass in both comparisons. For the wet-diameter comparison, we use versions
of all five AeroMELD decoders trained on wet-diameter targets.

After 12~h, the overall error is 4.1\% for WRF-AeroMELD and
12.2--21.5\% for the sectional and modal schemes. The latter errors are
3.0--5.2 times as large as WRF-AeroMELD's, similar to the range of
3.1--5.0 in the dry-diameter comparison. In wet diameter, WRF-AeroMELD
also has the lowest errors for both the number and speciated mass
distributions (Table~\ref{tab:size_distributions}).

The emission population illustrates why wet diameter is challenging for
the sectional schemes. Water accounts for about three quarters of its wet
mass, but individual particles differ in water content. MOSAIC groups
particles by dry diameter and TOMAS by dry mass. Both assign a single water
mass fraction to all particles in a bin. Thus, if fresh soot and a
water-rich particle have the same dry diameter and occupy the same bin,
the scheme assigns them the same wet diameter. Finer bins retain this
assumption of uniform composition within each bin.

\paragraph{Displacement in particle size.}
Equation~\eqref{eq:error} compares distributions bin by bin. Even a small
shift across a bin edge produces a deficit in one bin and an excess in the
next. We therefore also calculate the 1-Wasserstein distance in
$\log_{10}D$. After each distribution is normalized to unit total, this
distance measures the mean shift in log diameter needed to align the
predicted distribution with the reference. For speciated mass, we calculate
the distance separately for each species present in both distributions and
average the distances in proportion to their reference masses.

In dry diameter, TOMAS-40 has the smallest displacement for particle
number, and MOSAIC~8 and TOMAS-40 have the smallest displacements for
speciated mass. In wet diameter, WRF-AeroMELD has the smallest
displacement for both distributions (Table~\ref{tab:size_distributions}).

\begin{table}
\caption{Size-distribution errors after 12~h, averaged over mixed grid
cells. The wet-diameter errors use Equation~\eqref{eq:error}.
Wasserstein distances are in decades of particle diameter. One diagnostic
bin spans 0.25 decades. Mass denotes speciated mass. The lowest value in each column is bold.}
\label{tab:size_distributions}
\centering
\begin{tabular}{l rr rr rr}
\hline
 & \multicolumn{2}{c}{Wet-diameter}
 & \multicolumn{4}{c}{Wasserstein distance (decades)} \\
\cline{4-7}
\rule{0pt}{2.6ex} & \multicolumn{2}{c}{error (\%)}
 & \multicolumn{2}{c}{Dry diameter}
 & \multicolumn{2}{c}{Wet diameter} \\
\cline{2-3}\cline{4-5}\cline{6-7}
\rule{0pt}{2.6ex}Scheme & Number & Mass & Number & Mass & Number & Mass \\
\hline
WRF-AeroMELD & \textbf{6.1} & \textbf{9.4} & 0.025 & 0.025 & \textbf{0.021} & \textbf{0.045} \\
MOSAIC~8     & 28.0 & 29.7 & 0.090 & \textbf{0.017} & 0.141 & 0.115 \\
TOMAS-40     & 13.0 & 28.5 & \textbf{0.018} & 0.018 & 0.082 & 0.115 \\
MAM4         & 13.1 & 28.9 & 0.043 & 0.117 & 0.061 & 0.126 \\
MATRIX       & 18.6 & 23.7 & 0.108 & 0.098 & 0.125 & 0.110 \\
\hline
\end{tabular}
\end{table}

\paragraph{Reconstruction and averaging errors.}
Following \ref{app:attribution}, we calculate the aerosol
diagnostics separately for each represented source population and combine
them in the reference proportions. The reconstruction error compares these
combined diagnostics with the particle-resolved reference. The averaging
error compares them with the diagnostics obtained when we first mix the
represented source populations in the same proportions and then calculate
the diagnostics. The sum of the two errors bounds the directly measured
representation error. WRF-AeroMELD has the smallest error by both measures
(Table~\ref{tab:representation_errors}). Among the sectional and modal
schemes, TOMAS-40 has the smallest reconstruction error and MATRIX has the
smallest averaging error.

\begin{table}
\caption{Reconstruction, averaging, and directly measured representation
errors (\%) after 12~h, averaged over mixed grid cells. The size
distributions use dry diameter. WRF-AeroMELD results are for the
representative training seed. The lowest value in each column is bold.}
\label{tab:representation_errors}
\centering
\begin{tabular}{lrrr}
\hline
Scheme & Reconstruction & Averaging & Representation \\
\hline
WRF-AeroMELD & \textbf{3.6} & \textbf{2.2} & \textbf{3.6} \\
MOSAIC~8     & 14.2 & 5.0 & 18.8 \\
TOMAS-40     &  7.2 & 5.7 & 12.1 \\
MAM4         & 16.1 & 5.4 & 18.2 \\
MATRIX       & 11.3 & 3.3 & 11.8 \\
\hline
\end{tabular}
\end{table}

For the sectional and modal schemes, we report the transport error
directly. Separating it into fraction and off-mixture errors requires
estimating the source fractions from the transported tracers. When the
urban contribution is small, even a slight error in its fitted fraction
can appreciably change the predicted frozen fraction. For MOSAIC~8,
for example, the fit gives a fraction error of 1.5\% and an off-mixture
error of 1.4\%, although the directly measured transport error is only
0.4\%. The direct comparison avoids this sensitivity to the fitted
source fractions.

\section{Training and Testing the Decoders}
\label{app:training}

\paragraph{Diagnostics.} The training targets are computed from the particles
of each population: the activated fraction from each particle's critical
supersaturation by $\kappa$-K\"ohler theory at 287~K \cite{Riemer2010},
scattering and absorption by Mie theory with a core--shell treatment of black
carbon and dust \cite{Yao2022}, and immersion freezing with an active-site
parameterization \cite{Hoose2012,niemand2012particle,Schill2020,Tang2026}.
The decoder outputs the number size distribution and the speciated mass size
distribution in 15 bins of dry diameter up to 10~$\mu$m, the latter with
water, the activated fraction at 100
supersaturations from 0.1 to 10\%, the frozen fraction at 100 temperatures
from $-40$ to $0\,^{\circ}$C, and the scattering and absorption coefficients
per particle at eight wavelengths from 300 to 1000~nm.

\paragraph{Models.} All models share the encoder of
Equation~\eqref{eq:encoder}: a per-particle MLP (two hidden layers of 256) on
standardized square-root species masses, followed by a softplus, so that
$\zlat\ge0$, and the number-weighted mean, with nine latent variables. The
nonlinear decoders are MLPs with two hidden layers of 256. Every decoder has
nonnegative outputs, and builds the activated and frozen fractions as
cumulative sums of nonnegative increments, so that they are monotone.

\paragraph{Training.} The models are trained as variational autoencoders
\cite{Kingma2014}, with a standard normal prior and a Kullback--Leibler weight
of 0.03, using Adam \cite{kingma2014adam} on minibatches of 16 populations for
$10^5$ iterations; the learning rate rises linearly to $3\times10^{-3}$ over
the first 1000 iterations and then decays to $10^{-5}$ on a cosine schedule
(scaled by 0.3 for the nonlinear decoders' parameters and by 10 for the affine
decoder's), and gradient norms are clipped at 1. The transported latent vector
is the mean of the encoder's posterior. The loss compares the decoded and true
per-particle diagnostics: linearly for the CCN spectrum and the number size
distribution, by a Huber loss of the logarithmic difference for the optics and
the frozen fraction, and by the relative error of the test metric for the mass
distributions, with the activated- and frozen-fraction terms weighted by three.

\paragraph{Training on mixtures.} Training on mixtures adds the same loss on
mixtures of training populations: volume fractions are drawn from a uniform
Dirichlet distribution and converted to number fractions, the latent vector of
the mixture is the number-weighted average of Equation~\eqref{eq:mix}, and the
targets are the mixture's exact diagnostics. WRF-AeroMELD's decoder
(decoder~3 of \ref{app:decoders}) draws the number of populations uniformly
from $\{2,4,8,16\}$. Of the alternatives, decoder~2 (pairs) weights the
mixtures by 10 relative to the single populations, decoder~4 (up to 32) draws
from $\{2,4,8,16,32\}$, and decoder~5 (affine) is trained on pairs with weight 1.
Decoder~1 (no mixtures) is also trained without the Kullback--Leibler term.
With the term but without mixtures, the nonlinear decoder failed to train. Its
overall error on single test populations was 82--91\% (four seeds), compared
with 4.1--6.0\% for decoders~2--4. We therefore tested the term
on decoder~3 instead, training it without the term (ten seeds;
\ref{app:decoders}).

\paragraph{Test mixtures.} The test mixtures of $K$ populations draw their
volume fractions from a uniform Dirichlet distribution, converted to number
fractions as in training; 2000 mixtures are drawn for each $K$ and seed, the
same for every decoder.

\section*{Open Research Section}
The WRF-AeroMELD implementation with per-tracer latent advection, a
modification of WRF version~4.7.1 \cite{Skamarock2019}, is archived as
\citeA{wrfaeromeld2026} (release \texttt{wrf-aeromeld-v1.0} of the
\texttt{latent\_per\_tracer} branch of \url{https://github.com/compdyn/WRF}).
The library of PartMC-MOSAIC scenarios used for training and for the source
populations is available from the Illinois Data Bank \cite{Gasparik2020data},
and the NAM analyses that drive WRF from the NSF NCAR Geoscience Data Exchange
\cite{NCEP2015NAM}. The particle dataset was built from the library with the
\texttt{dataset\_build} scripts of partnn, archived as \citeA{partnn2026}
(release \texttt{aeromeld-wrf-v1.0} of \url{https://github.com/ehsansaleh/partnn}).
The training and evaluation of the models, the offline transport, the sectional
and modal schemes, the analysis and the figure scripts are archived as
\citeA{aeromeldcode2026} (release \texttt{v1.0} of
\url{https://github.com/compdyn/aeromeld-wrf-advection}) under the MIT license,
and the training targets, the trained models, the WRF input files, the transport results and the
processed data as \citeA{aeromelddata2026} under the CC BY 4.0 license. PartMC
\cite{partmc2025} and PyPartMC \cite{pypartmc2025} were used to simulate the
library and to compute diagnostics of the source populations. The analysis used
NumPy \cite{Harris2020}, SciPy \cite{Virtanen2020}, PyTorch \cite{Paszke2019}
and Matplotlib \cite{Hunter2007}.

\section*{AI Tools Disclosure}
The code written for this study was implemented by Anthropic's Claude
(Claude Opus~5.5, through Claude Code): the per-tracer latent advection in WRF,
the offline transport, the training and evaluation of the models, and the
analysis and figures. Claude also helped write this manuscript. The authors
checked all of the code, results and text, and take full responsibility for
the content.

\section*{Conflict of Interest}
The authors declare no conflicts of interest relevant to this study.

\acknowledgments
This work was supported by the U.S. Department of Energy, Office of Science,
Office of Biological and Environmental Research under Award Number
DE-SC0022130.

\bibliography{refs_wrf_ml}

\end{document}